%% file: main.tex
\documentclass[nonacm]{acmart}

\setcopyright{acmcopyright}
\copyrightyear{2024}
\acmYear{2024}
\acmDOI{XXXXXXX.XXXXXXX}

\acmConference[CHI `24]{2024 CHI Conference on Human Factors in Computing Systems}{May 11--16, 2024}{Honolulu, HI}

\usepackage{graphicx}
\usepackage{multirow}
\usepackage{subfig}
\usepackage{hyperref}
\usepackage{color}
\usepackage[normalem]{ulem}
\usepackage{tabularray}

\newcommand\blfootnote[1]{%
  \begingroup
  \renewcommand\thefootnote{}\footnote{#1}%
  \addtocounter{footnote}{-1}%
  \endgroup
}

\title[The Cadaver in the Machine]{\emph{The Cadaver in the Machine}: The Social Practices of Measurement and Validation in Motion Capture Technology}

\author{Emma Harvey}
\affiliation{%
  \institution{Cornell University}
  \city{Ithaca}
  \state{New York}
  \country{USA}}
\email{evh29@cornell.edu}

\author{Hauke Sandhaus}
\affiliation{%
  \institution{Cornell University}
  \city{New York}
  \state{New York}
  \country{USA}
}
\author{Abigail Z.\ Jacobs}
\affiliation{%
 \institution{University of Michigan}
 \city{Ann Arbor}
 \state{Michigan}
 \country{USA}}
\authornote{All three authors are Principal Investigators and contributed equally to leading the paper, research design, and analysis.}

\author{Emanuel Moss}
\affiliation{%
  \institution{Intel}
  \city{New York}
  \state{New York}
  \country{USA}}
\authornotemark[1]

\author{Mona Sloane}
\affiliation{%
  \institution{University of Virginia}
  \city{Charlottesville}
  \state{Virginia}
  \country{USA}}
\authornotemark[1]

\begin{document}

\begin{abstract}
Motion capture systems, used across various domains, make body representations concrete through technical processes. We argue that the measurement of bodies and the validation of measurements for motion capture systems can be understood as \textit{social practices}. By analyzing the findings of a systematic literature review (N=278) through the lens of social practice theory, we show how these practices, and their varying attention to errors, become ingrained in motion capture design and innovation over time. Moreover, we show how contemporary motion capture systems perpetuate assumptions about human bodies and their movements. We suggest that social practices of measurement and validation are ubiquitous in the development of data- and sensor-driven systems more broadly, and provide this work as a basis for investigating hidden design assumptions and their potential negative consequences in human-computer interaction.\blfootnote{\copyright 2024. Copyright held by the authors. This is the authors' version of the work. The version of record will be published in \textit{2024 ACM CHI Conference on Human Factors in Computing Systems (CHI ’24)}.}
\looseness=-1

\end{abstract}

\ccsdesc[500]{General and reference~Validation}
\ccsdesc[500]{General and reference~Measurement}
\ccsdesc[300]{General and reference~Evaluation}
\ccsdesc[500]{Human-centered computing~Human computer interaction (HCI)}
\ccsdesc[300]{Social and professional topics~Socio-technical systems}
\ccsdesc[300]{Hardware~Hardware validation}

\keywords{motion capture, measurement, validation, social practices, anthropometry}

\maketitle

\input{sections/01introduction}
\input{sections/02socialpractices}
\input{sections/03method}
\input{sections/04findings}
\input{sections/06discussion}

\begin{acks}
This work was supported by funding from the Notre Dame-IBM Technology Ethics Lab. We thank Todd Bryant, Luke DuBois, Kaustav Sarkar, and Harsh Palan for their invaluable explanations of motion capture processes.
\end{acks}

\bibliographystyle{ACM-Reference-Format}
\bibliography{references}

\end{document}

%% file: sections/01introduction.tex
\section{Introduction}\label{sec:intro}

Motion capture systems are widely used to measure bodies and their movement. Representations of bodies inferred by motion capture systems are used in a range of fields including entertainment, manufacturing, medicine, sports, and robotics \cite{nogueira_motion_2012, zhou_human_2008,he_survey_2019,song_HumanPoseEstimation_2021,menolotto_motion_2020, bregler_MotionCaptureTechnology_2007,magnenat-thalmann_ModellingMotionCapture_1998}. Applications include generic settings like gesture detection, tangible interfaces, and pose estimation, as well as more sensitive (and fraught) applications like health diagnosis, surveillance, and emotion or gender detection \cite{flux_human_2020,cunha_neurokinect_2016,das_quantitative_2011,smith_gender_2002, deligianni_emotions_2019,barnachon_ongoing_2014,boualia_pose-based_2019}.
\looseness=-1

Motion capture systems and their applications are based on assumptions about how and what we can learn about human bodies and their movement. These assumptions require interrogation, as they underwrite the uses to which systems are put and shape the material effects of these systems in the world. We draw special attention to how the ubiquitous and concomitant activities of \textit{measurement} and \textit{validation} co-develop over time and  carry forward such assumptions. We show that errors play a central role in identifying hidden, historically layered, and contingent assumptions in motion capture technology. By analyzing measurement and validation as \emph{social practices}, we show that social practice theory offers a crucial lens into the ways in which these assumptions---and their potential harms---become hidden and entrenched. 

\subsection{Preliminaries: Inferring skeletons, body segments, and their motion.}\label{subsec:prelim-skeletons} 
Motion capture technologies infer the position, rotation and translation of bodies in space from data collected by sensors. Sensors may be placed directly on the surface of the body (e.g., inertial measurement units (IMU) or accelerometers) or may be external cameras that sense markers placed on the body (e.g., optoelectronic systems) or sense the surface of the body directly (e.g., depth cameras). Data from these sensors are used to infer features and movement of bodies and their components (limbs, torso, head, feet, and hands).
This requires statistical inference about the relationship between bodies' surfaces, skeletal elements, and non-skeletal elements (e.g., soft tissue, clothing) \cite{nogueira_motion_2012}. These inferences rely on the results of prior studies, which were conducted to establish models of the human body and movement through practices of measurement and validation.

Several key terms emerge. While these terms shift meanings and use over time, we introduce them here for the reader to better clarify our findings in the following sections.
\emph{Skeletons} are typically inferred from motion capture methods, and may refer to representations of the body's skeletal structure or to the generic body representation output of motion capture systems. Inferring the motions (\emph{kinematics}) and forces (\emph{dynamics}) of these (also inferred) skeletons relies on simplified models of bodies.\footnote{See Vaughan et al.\ \cite{vaughan_dynamics_1992} for an accessible overview of dynamics for human gait analysis.} Consider estimating the amount of force exerted at a joint, such as a knee. This is done through stylized models of the kinematics and dynamics of components of the body and how they relate to each other. First, the \textit{body segments} corresponding to the thigh and calf must be identified, and their motions measured. Then, the amount of force theoretically required to cause the observed motion must be calculated. That calculation itself requires knowledge of a subject's \textit{body segment parameters} (BSPs): characteristics including mass and center of gravity that are not directly observable without invasive procedures up to and including dismemberment. As a result, BSPs must also be inferred from formulas based on an individual's \textit{anthropometric measurements} \cite{vaughan_dynamics_1992}.

Through a lens of social practice theory, we show these elements of motion capture rest on normative assumptions about who should be measured to understand the human body: those who are male, white, `able-bodied,' and of unremarkable weight. We further show that these assumptions continue to be implicit in modern motion capture due to practices of validation that entrench, rather than challenge, them. We note that physiological and anthropometric measurement and representation have been foundational to the field of HCI and human factors, reflecting the ways that physical interactions and arrangements of bodies constrain and enable computational interactions \cite{hewett_acm_1992}. As the body becomes a site of interface and interaction with computer systems \cite{dourish_where_2001} through VR, AR, and XR, the assumptions that underpin its symbolic representation in those systems will have a material effect on such interactions \cite{klemmerHowBodiesMatter2006,gerling_critical_2021}.

Throughout this work we attend to the shifting ways in which motion capture design engages with \emph{measurement}, i.e., the practices concerned with the standardization of generating quantitative knowledge (such as of body features), and \emph{validation}, concerned with the conformity to, or violation of, standards for generating quantitative knowledge based on previously established measurement practices. 
A key theme is concerned with types of \emph{errors}. Measurement practices center around known sources of error, and validation practices center around the identification and management of those errors. While our focus is a critical unearthing of the  social practices of measurement and validation in motion capture, 
adjacent literatures offer a critical lens to examine the representations of bodies in extended reality and CHI more broadly \citep[e.g.,][]{gerling_critical_2021,sum2022dreaming,williams2021articulations}; to empirically study knowledge infrastructures \cite{karasti_studying_2018,star_ethnography_1999,simonsen_infrastructuring_2020,muller_designing_2021}; to understand how assumptions are encumbered and carried forward by communities in data-driven systems \cite{passi_data_2017,passi_trust_2018,pasquetto_uses_2019,jaton_we_2017}, and the implications and politics thereof \cite{pine_politics_2015, pierce_sensor_2020}; and to measurement, validation,\footnote{We also note specific engagement with the \emph{validity} of motion capture systems for the design of HCI-related tasks and human factors more broadly. Bachynskyi et al.\ \cite{bachynskyi_is_2014}, for instance, explores the validity of motion capture systems for HCI-related tasks, such as playing a dance game, typing, and steering a plane. Bachynskyi et al.\ argued that motion capture for HCI must be broadly applicable to people of varying demographic and physical characteristics, and is more interested in the effects of motion (e.g., muscle stress or fatigue) than precisely quantifying the motions themselves. Crucially, they noted that ``biomechanical models cannot be considered as `valid in general,' but they have to be validated for each application or task" \cite{bachynskyi_is_2014, barocas_hidden_2019}. 
In contrast we undertake a systematic review to understand what is meant by ``validation''---and what that implies for HCI research.} 
and error \citep[e.g.,][]{jacobs_measurement_2021,coston2023validity,lin2023bias}.

\subsection{Contributions.} 
This paper offers a detailed case study of the emergence and stabilization of the assumptions built into motion capture technologies. Additionally, this paper also contributes an articulation of how social practice theory \cite{shove_dynamics_2012} can be deployed in the analysis of human-computer interactions. It does so by applying a social practice theory lens to the historical development of measurement and validation practices in motion capture technology design and innovation. This
approach focuses on \emph{practices} as the unit of inquiry and attends to how the elements of social practice, i.e. \emph{meanings, materials}, and \emph{competences}, link up to stabilize a social practice across time, space, and contexts. It allows analysts to attend to the role assumptions play in stabilizing measurement and validation practices in motion capture from its inception to today. 
\looseness=-1

We argue that it is essential to examine the history of motion capture, and specifically of measurement and validation in motion capture technology, to reveal the dynamic interaction between assumptions and technology design. To those ends, we articulate how assumptions are embedded in all sociotechnical systems and introduce the analytical lens of \emph{social practices} of measurement and validation in \S\ref{sec:socialpratices}. We argue that historical analysis is key for this approach and describe our systematic literature review of motion capture design, measurement, and validation from 1930s to today, in \S\ref{sec:methods-litreview}. In \S\ref{sec:findings-errors}, we offer a history of measurement practices surfaced by our review and characterize how different types of errors reveal stabilizing practices in each historical era. We uncovered three distinct phases of historical development in which measurement practices concretely stabilize as distinct types of errors take center stage. Similarly, we describe when and how validation practices stabilized (\S\ref{sec:findings-validation}) and where these measurement and validation practices encode assumptions that underpin motion capture technology and continue to underpin the basic functioning of the technology today. 
Finally, in \S\ref{sec:discussion} we show how this type of analysis of social practices reveals the hidden stakes of related artificial intelligence-based technologies that inherit the assumptions of their contexts. Such a perspective thus makes possible meaningful holistic AI audits for accountability \cite{sloane_silicon_2022}.  
\looseness=-1

%% file: sections/02socialpractices.tex
\section{Social practices of measurement and validation}\label{sec:socialpratices}
All sociotechnical systems are based on assumptions: about things, about people, and about how the world works or ought to work. Assumptions are not inherently bad; they guarantee a basis upon which people and systems can interact. But assumptions baked into sociotechnical systems can cause harms. This is particularly the case when they enshrine stereotypes and prejudices that cause \textit{representational} harms (for example, sexualizing Black women and girls \cite{Noble_2018}, \textit{allocative} harms (for example, keeping poor people from accessing resources \cite{eubanks_automating_2017, barocas_fairness_2023}), or \textit{bodily} harms (for example, through "race-norming" in diagnostics \cite{braun_breathing_2014, gasquoine_race-norming_2009}). Even though they are the root cause of these harms, assumptions can be hard to identify and exorcise. They are the background knowledge, taken-for-granted approaches, and unreflected-upon aspects of an activity or understanding. Assumptions remain unseen.

In the following we reveal how assumptions become embedded in sociotechnical systems using the analytic lens of social practice theory, focusing on two social practices in particular:  measurement and validation. We argue that paying attention to how the elements of measurement and validation practices in motion capture combine and stabilize can help excavate harmful assumptions. Specifically, we show how attending to errors---the targets of specific improvement at the time of development---within the practices of measurement and validation offers a useful point of entry into understanding these practices and assumptions.

\paragraph{All sociotechnical systems rest on assumptions}
Assumptions exist for good reasons. We must assume that certain things are true in order to stabilize provisional knowledge into facts through the accumulation of supporting evidence, or to alter our knowledge in the face of contradictory evidence \cite{gitelman_data_2013}. Multiple narratives of the history and philosophy of science illustrate how the production of scientific knowledge depends on assumptions about the natural world that, once stabilized, can be set aside from the focus of subsequent experiments \cite{hacking_representing_1983, latour_science_2003, shapin_leviathan_1985}, at least until they must be revisited in the face of confounding evidence. Nuclear models have been revised as assumptions about subatomic particles’ properties were updated \cite{barad_meeting_2007}, as have assumptions about immunity and illness \cite{martin_flexible_1999}. These revisitations demonstrate that even as prior assumptions become embedded in what we know about the world and the sociotechnical systems we build to operate on it, they remain latent artifacts of once-provisional knowledge now taken as given. 
\looseness=-1

Because they remain as the ground upon which knowledge and technologies are built, unpacking assumptions is necessary to understand how they have shaped the practices around testing and verifying the science and technology that was built upon them.  Unpacking assumptions is particularly important as technologies proliferate to interact with broader populations and changeable environments to ensure that they remain sound and do not cause harm, given their scope of use. Particularly with data-centric technologies, where data collected from within one social milieu might be used to develop products for different or broader social settings \cite{koh_wilds_2021}, it also becomes necessary to reveal and probe the assumptions that stabilize the `ground truth’ against which the system is validated. Finally, unpacking assumptions must then account for the histories through which systems were developed \cite{edwards_knowledge_2013}, and the stabilizing of what is taken to be `ground truth’ within them.\looseness=-1

\input{figures/elements}

\paragraph{Analytical lens: social practices as the unit of inquiry} 
The analytical frame we deploy is informed by social practice theory \cite{shove_dynamics_2012, shove_design_2007}. This approach is motivated by a well-established tradition of examining ``the social'' within technology by way of practices, most notably led by figures like Donna Haraway \cite{haraway_situated_1988} and Lucy Suchman \cite{suchman_reconstructing_1999}. Rather than taking an object, an individual, or a system as a unit of inquiry, social practice theory focuses on behaviors and activities that span across communities and across time. The emphasis is on understanding how stability and change emerge in the process of the ``elements” of a social practice linking up or breaking apart in specific cultural contexts: materials, competences, and meanings \cite{shove_design_2007, shove_dynamics_2012}, as illustrated by Figure \ref{elements_fig}. 
\textit{Materials} are objects, technologies, infrastructures, tools, hardware, bodies, and stuff; \textit{competences} are physical and mental skills, multiple forms of shared understanding, and practical knowledgeability; and \textit{meanings} are the social and symbolic significance, image, convention, aspiration, belief or identity at play  \cite{shove_dynamics_2012, higginson_diagramming_2015}.  

For example, the practice of ``running” is composed of the \textit{materials}  human bodies, running clothes and shoes, perhaps headphones and a smartphone; the \textit{competence}  running or jogging for a sustained period of time; and the \textit{meaning} of moving one's body for the sake of health or athletics. These elements link up and stabilize the practice of ``running” writ large by way of being repeated by many different people at different times and in different places. In other words, a critical mass of these individualized repetitions constitute a ``social practice'' that takes on its own career trajectory. In another example, ``voguing'' only stabilized as its present day dance style practiced by many people in many different places once it went mainstream in the 1990s following a feature in documentaries and a Madonna music video, even though it had been around for decades before that \cite{becquer1991}, originally practiced by Black and Latinx queer communities in New York City in gatherings called ``balls'' \cite{art_of_voguing, history_of_voguing}. The practice of voguing began to take on its own career, as depicted in Figure~\ref{voguing}.

\input{figures/voguing}

The elements of social practices shape each other over time. That is, materials affect meanings and meanings affect competences, and so on;  different social practices can share one or more elements, creating an evolving nexus of practices. For example, the practice of measuring is conditioned on the availability of the competence of applying an agreed-upon metric, for example to a body. That competence is an element shared with many different practices, ranging from XR system design to medical diagnostics.

All practices and sociotechnical practices, like motion capture, inherit assumptions baked into earlier versions of elements. \textbf{These assumptions \textit{spread} and \textit{solidify} as the practice evolves.} As we demonstrate in the following, a social practice theory approach is useful for understanding how the practice of measuring bodies first stabilized based on a certain set of a assumptions in very specific contexts, and how these continue to be stabilized in machine-driven measurements of bodies in motion capture systems. The basis for this approach is paying heightened attention to the assumptions that continue to flow from the first instances of measuring bodies for motion capture, in other words: how assumptions inevitably continue to stabilize alongside that (now machine-driven) practice.\looseness=-1

\paragraph{Social practices reveal assumptions in technologies.} 
A social practice theory approach is committed to understanding established patterns of action and social structure, including in their material dimension. This, explicitly, includes dominant (cultural) assumptions \cite{shove_dynamics_2012}. Assumptions often both motivate the emergence or profound reconfiguration of a social practice in the first place, and they are profoundly related to the element `meaning.' For example, the assumption that power ought to be linked to masculinity led, according to the technology historian Mar Hicks \cite{hicks_programmed_2018}, to the downfall of the British computing industry after World War II: as soon as computers, originally associated with secretarial work and femininity and little power, became seen as central to power within organizations and society at large, highly skilled woman computer experts were replaced by low-skill male managers. The practice of computer work profoundly changed based on assumptions around power and masculinity and re-stabilized with a new meaning that was `gender-flipped.' Assumptions that we carry today around innate ability to do computer work stem, in some parts, from that particular moment in time. Therefore, assumptions not only motivate (changes of) social practice, but are also the result.\looseness=-1 

As we posit here, revealing assumptions and how they get baked into sociotechnical systems increasingly plays a role in understanding the harms these systems can cause, particularly along the fault lines of intersectional identifiers such as race, gender, and class \cite{bolukbasi_man_2016, buolamwini_gender_2018, obermeyer_dissecting_2019, stypinska_ageism_2021, eubanks_automating_2017}. For example, the assumption that Black people's lung capacity is `naturally' lower led to `race-norming' in spirometer technology, causing Black workers having to demonstrate a higher degree of impairment than white peers before receiving medical treatment \cite{braun_breathing_2014}. 
A similar example is `race-norming' in neuropsychological tests which, until recently, were in part used in settlement payout processes in traumatic brain injury cases in the NFL \cite{hobson_how_2021}. Here, too, the underpinning assumption was Black individuals generally score lower in neuropsychological tests \cite{gasquoine_race-norming_2009}, which led to automatic statistical adjustments of test scores with the result of injured Black players having to demonstrate a much higher degree of injury (i.e. much lower test score) than white players-- a practice that was recently ended \cite{belson_changes_2022}. 
A social practice theory approach allows us to not only identify \textit{what} assumptions are at play, but also to examine \textit{how} assumptions become stabilized and infrastructural -- an examination that is particularly important for understanding how assumptions become an implicit part of technology design and human-computer interaction.\looseness=-1

\paragraph{Principal sociotechnical practices of motion capture: measurement and validation.} 
Social practices overlap and are composed of constituent practices. `Motion capture' is a sociotechnical practice that is composed of multiple other practices, including `measurement' and `validation' as well as others like `hardware design', `innovation,' and `documentation'. The practices of measurement and validation, however, take on specific significance in motion capture: they are foundational to both the technology and the practice, as we will demonstrate in subsequent sections. 

\input{figures/elements_measurement} 

The \textit{materials} of measurement practice in motion capture are cameras, bodies, markers, bodysuits, software, agreed-upon metrics, recording studios, room calibration tools; the \textit{competence} is applying an agreed-upon metric to bodies in space using tools; and the \textit{meaning} is creating representations of bodies in space, i.e., the ability to transform measures into representations that are useful for the applications of motion capture technology (Figure \ref{elements_measurement}). These elements link up into measurement practice which, in motion capture, is concerned with the standardization of generating quantitative knowledge (such as of body features) i.e., quantifying known sources of error. 

The \textit{materials} of validation practice in motion capture are statistical models of bodies and movement, measurements of bodies; the \textit{competence} is applying an agreed-upon framework for evaluating a specific set of measurements; and the \textit{meaning} is improving motion capture technology, i.e., evaluating a set of measurements to ensure they are a reasonable proxy for the real world (Figure \ref{elements_validation}). These elements link up into validation practice which, in motion capture, is concerned with the conformity to or violation of standards for generating quantitative knowledge (based on previously established measurement practices), i.e., identifying and managing errors. 

\input{figures/elements_validation} 

Measurement and validation practices in motion capture converge in the element `meaning': both practices are meaningful as part of design and innovation in motion capture technology. In both practices, assumptions most prominently shore up in errors, because what is considered an `error' at any point is a locus of attention to what can be controlled and minimized (through measurement) and identified and managed (through validation).
\looseness=-1

\paragraph{Errors offer key insights into motion capture practices.}
Critical archives and science and technology scholars have described how {errors} reveal departures of technical models from social expectations \cite{amrute_techno-ethics_2019, lin2023bias, hutter_organizational_2005}. 
Similarly, recent calls for human-centered design for machine learning point to errors as an opportunity to re-imagine those systems \cite{kogan_mapping_2020, chancellor_toward_2023}. This work seeks to reframe data creation and analysis as a site of critical and of practical intervention---where errors often point to such a site \cite{muller_how_2019,passi_data_2017,passi_trust_2018, cabrera_what_2023, miceli_studying_2022, passi_making_2020}. Fundamentally, the phenomena that practitioners attend to \emph{as} errors point to the aspects of their practices of measurement and validation that are of active and ongoing concern.  

Lin and Jackson \cite{lin2023bias} examine the ``practice of errors'' shared among machine learning practitioners, calling attention to the competences and skills involved in error management, the potential of errors for new discovery, and the fact that errors reveal hidden social structures and arrangements. They propose that errors always expose infrastructural assumptions, and that they have the capacity to create new or recombine social arrangements. 
Lin and Jackson's work converges with our social practice theory approach in that it is explicit about competences and the material context of the work of domain experts in technology design, and in that it sheds light on meaning-making across individuals and communities of practice. Most importantly, Lin and Jackson's empirical study underlines the significance of errors for understanding the stabilization and continuous reconfiguration of measurement and validation practices, because it demonstrates that errors can reveal existing social practices by highlighting elements or their relations, that errors can lead to new social practices via the incorporation of new elements, and that errors can trigger the re-configuring of social practices. 

\input{figures/error_lens}

Taken at a moment in time, and within the moment of practice stabilization, errors thus serve as an important site for understanding social practices and their elements. Different types of errors documented in different moments in time index the agreements and concerns that accompany specific sets of assumptions about what is relevant to the system. Therefore, it is evident that errors and the documentation of their evolution should be a key focus when working to understand stabilization and change in measurement practice and validation for motion capture technology.

%% file: figures/elements.tex
\begin{figure*}[h]
\centering
{\includegraphics[width=0.9\textwidth]{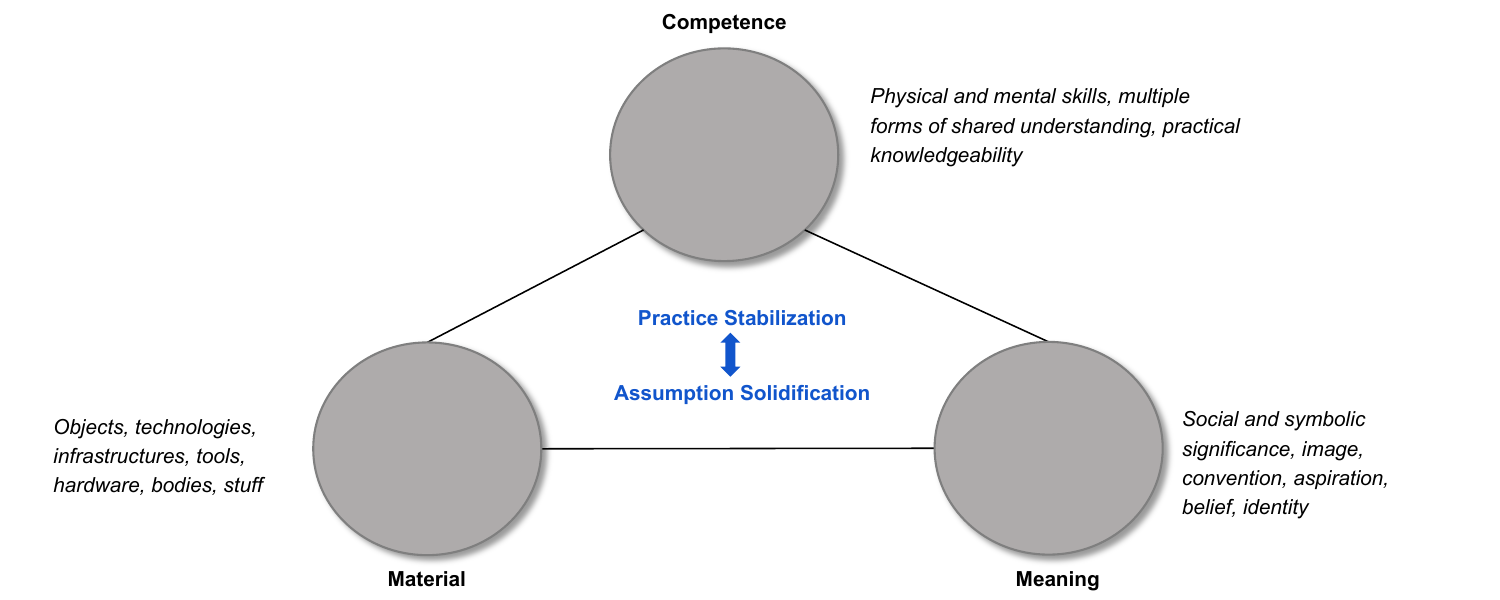}}
\hspace{0.1\textwidth}
\caption{\begin{small} 
\textbf{The elements of social practices.} Sociotechnical practices, like motion capture, inherit assumptions that are baked into their elements, including earlier versions of those elements. Assumptions that give rise to the practice in the first place also spread by way of those elements being combined and recombined. As practices evolve and stabilize, the assumptions that undergird them solidify.
\end{small}
\label{elements_fig}}
\Description{A conceptual diagram with three large, grey nodes connected by lines, forming a triangle. At the top node, labeled 'Competence', there is text that reads 'Physical and mental skills, multiple forms of shared understanding, practical knowledgeability'. The bottom left node, labeled 'Material', contains 'Objects, technologies, infrastructures, tools, hardware, bodies, stuff'. The bottom right node, labeled 'Meaning', includes 'Social and symbolic significance, image, convention, aspiration, belief, identity'. In the center of the triangle, there is a bidirectional arrow between two terms: 'Practice Stabilization' and 'Assumption Solidification'. This represents the dynamic interplay between material, competence, and meaning in social practices.
}
\end{figure*}

%% file: figures/voguing.tex
\begin{figure*}[h]
\centering
{\includegraphics[width=0.9\textwidth]{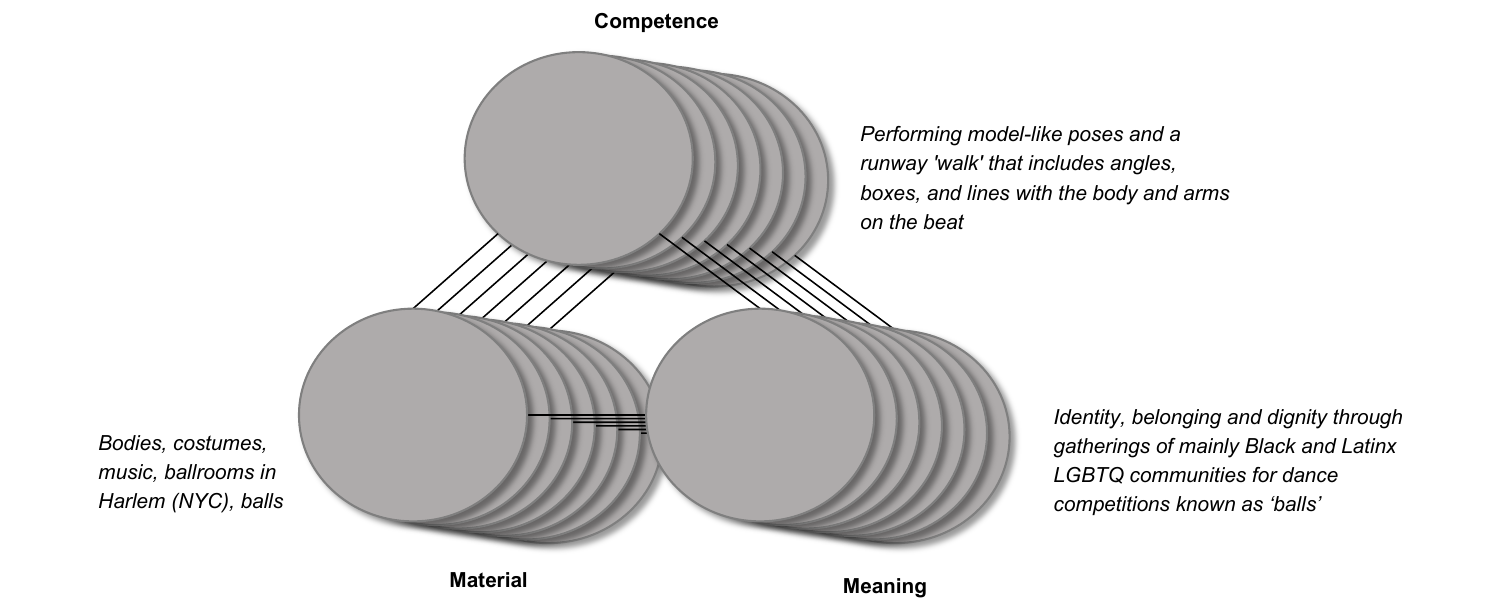}}
\hspace{0.1\textwidth}
\caption{\begin{small} 
Stabilizing “voguing” as a distinct dance form 1960s-1980s: critical mass of repetitions.
\end{small}
\label{voguing}}
\Description{A diagram similar to the previous figure but layered multiple times illustrating the elements of the dance form "voguing." Three overlapping ovals represent the components 'Competence,' 'Material,' and 'Meaning.' 'Competence' includes performing model-like poses and a runway walk that includes angles, boxes, and lines with the body and arms on the beat. 'Material' encompasses bodies, costumes, music, ballrooms in Harlem (NYC), and balls. 'Meaning' refers to identity, belonging, and dignity through gatherings of mainly Black and Latinx LGBTQ communities for dance competitions known as 'balls.' These elements overlap to depict how voguing is a complex, culturally rich dance form with deep historical roots and connections to specific communities.}
\end{figure*}

%% file: figures/elements_measurement.tex
\begin{figure*}[h]
\centering
{\includegraphics[width=0.9\textwidth]{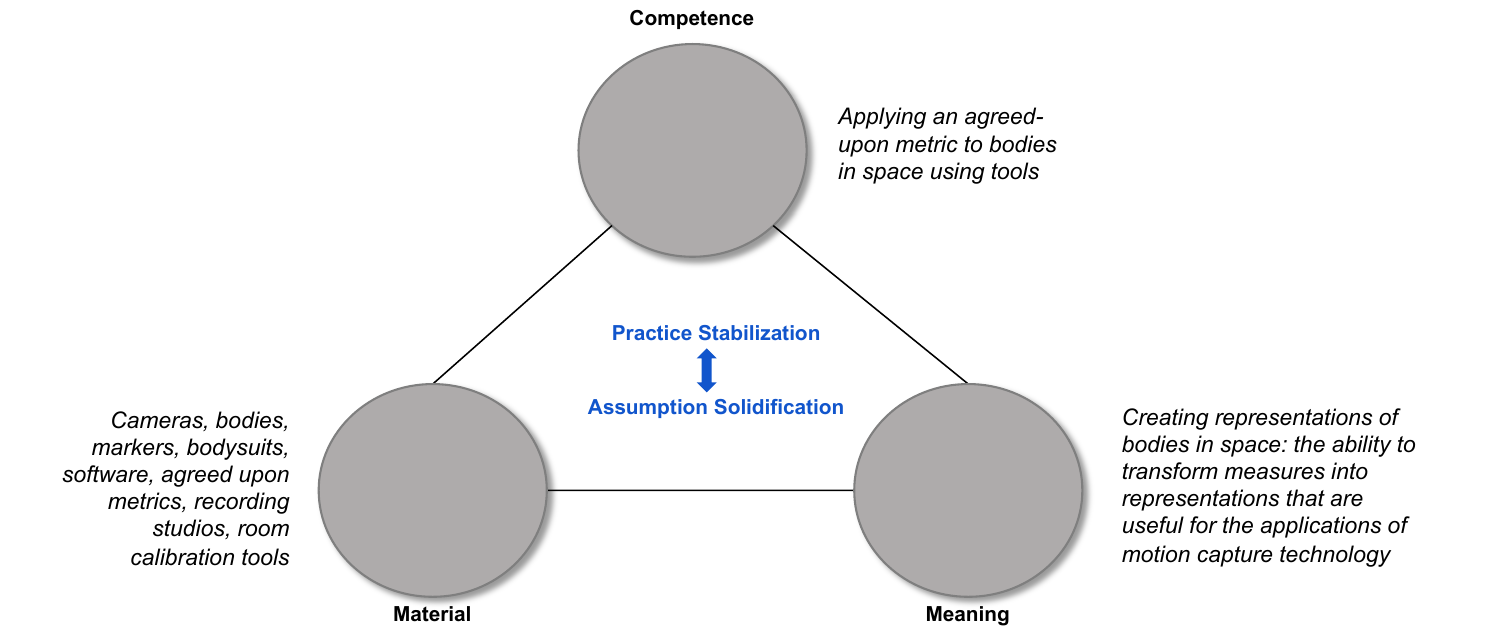}}
\hspace{0.1\textwidth}
\caption{\begin{small} 
The elements of \textit{measurement} practice for motion capture.
\end{small}
\label{elements_measurement}}
\Description{A diagram with three interconnected circles. The top circle, labeled 'Competence,' details 'Applying an agreed-upon metric to bodies in space using tools.' The bottom left circle, labeled 'Material,' lists 'Cameras, bodies, markers, bodysuits, software, agreed upon metrics, recording studios, room calibration tools.' The bottom right circle, labeled 'Meaning,' describes 'Creating representations of bodies in space: the ability to transform measures into representations that are useful for the applications of motion capture technology.' In the center, where all three circles intersect, are the words 'Practice Stabilization' and 'Assumption Solidification,' implying the foundational role of these concepts in the practice of motion capture measurement.}
\end{figure*}

%% file: figures/elements_validation.tex
\begin{figure*}[h]
\centering
{\includegraphics[width=0.9\textwidth]{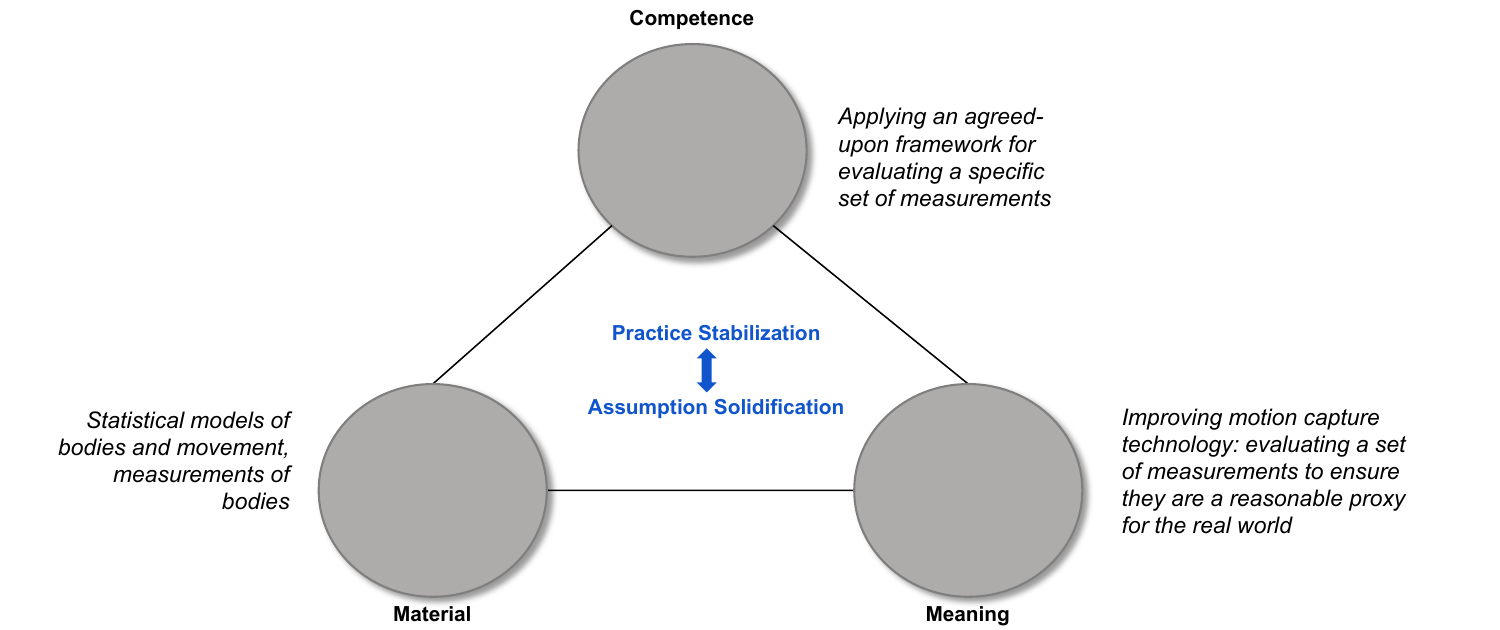}}
\hspace{0.1\textwidth}
\caption{\begin{small} 
The elements of \textit{validation} practice for motion capture.
\end{small}
\label{elements_validation}}
\Description{The diagram features three interconnected circles representing different aspects of motion capture validation practice. The top circle, labeled 'Competence,' includes the text 'Applying an agreed-upon framework for evaluating a specific set of measurements.' The bottom left circle, labeled 'Material,' lists 'Statistical models of bodies and movement, measurements of bodies.' The bottom right circle, labeled 'Meaning,' refers to 'Improving motion capture technology: evaluating a set of measurements to ensure they are a reasonable proxy for the real world.' At the center of the interconnected circles is 'Practice Stabilization' above 'Assumption Solidification,' suggesting these are central to the process of validating motion capture practices.}
\end{figure*}

%% file: figures/error_lens.tex
\begin{figure*}[h]
\centering
{\includegraphics[width=0.9\textwidth]{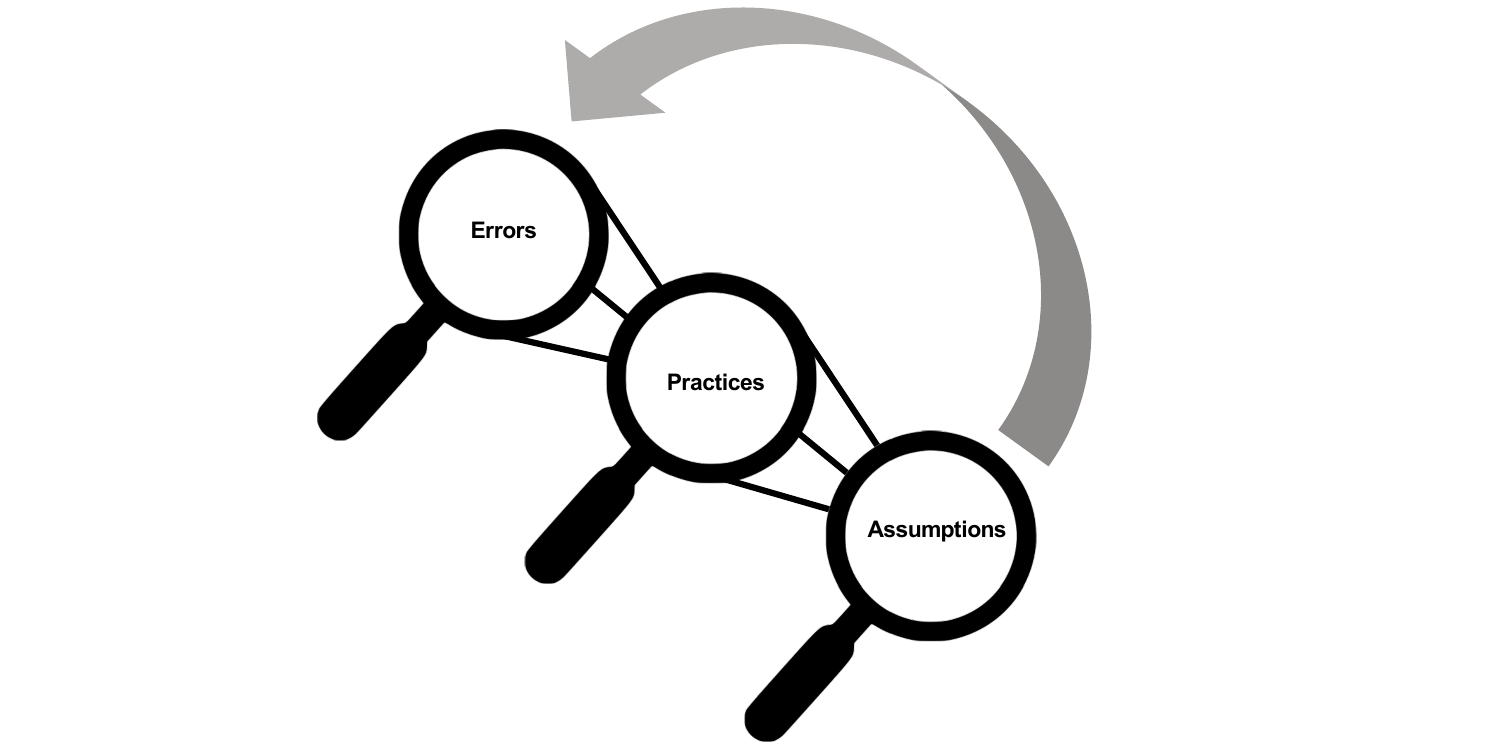}}
\hspace{-0.1\textwidth}
\caption{\begin{small} 
Analytical lens: errors provide insight into practices stabilization and assumptions solidification.
\end{small}
\label{error_lens}}
\Description{A diagram featuring a cyclical process with three magnifying glasses interconnected by arrows. Each magnifying glass focuses on a word: the first on "Errors," the second on "Practices," and the third on "Assumptions." The arrows indicate a directional flow from errors to practices, practices to assumptions and a large arrow pointing back to errors.}
\end{figure*}

%% file: sections/03method.tex
\section{Methods and Conceptual Analysis}
\label{sec:methods-litreview}

We apply the analytical lens of social practice theory to the motion capture literature on measurement and validation. Because we are interested in a historical view of the social practices of measurement and validation as they have been shaped and reshaped within motion capture, a systematic literature reviews offer a rigorous approach to uncover elements of social practices over time. We focus on the measurement of errors specifically as a lens through which to examine the assumptions inherent in the practices of measurement and validation more broadly.
To those ends, we conducted a systematic literature review of measurement, validation, and errors in anthropometry (referring to the measurement of the human body) and motion capture (referring to tracking technologies of 3D bodies in space). We characterize our findings in \S\ref{sec:findings-errors} and \S\ref{sec:findings-validation}.
\looseness=-1

\subsection{Empirical Studies of Social Practices}
Any empirical examination of the dynamics of a social practice is specific to that practice, its elements, and related practices. Thus uncovering assumptions in the social practices of measurement and validation for motion capture requires a practice-specific approach. Key here are three concerns: the process of standardization of measurements, the process of validation, and the significance of errors as insight into these social practices.\looseness=-1

Social practice scholars are typically concerned with observing and analyzing practices in the field by way of ethnography, for example, understanding the practices of technology use in the home \cite{zotero-4043}. At the same time, however, a social practice lens explicitly encompasses a view for the historic development of a practice and its elements. This is because of its commitment to understanding the moments in which stability and change happen in a social practice: both are staked on a relationship with history and context. Materials and objects, for example, are always newer versions of what was before. (The modern car, for example, is still an updated version of a carriage \cite{shove_dynamics_2012}.) This study is particularly interested in these historic continuities as it focuses on tracing the emergence and stabilization of assumptions in practices of measurement and validation in motion capture technology. A systematic literature review is one appropriate method for this tracing exercise, since the literature is where measurement standards and validation practices are documented. A systematic literature review is, therefore, well within the scope of a social practice theory approach.\looseness=-1 

\subsection{Systematic Literature Review}

In accordance with prior HCI literature \citep[e.g.,][]{bandy_problematic_2021, zhang_what_2023}, we relied on the Preferred Reporting Items for Systematic Reviews and Meta-Analyses (PRISMA) to guide our empirical data collection process \cite{PRISMA}. As such, we (1) developed a search strategy and conducted a keyword search of relevant databases, (2) screened the results of our keyword search for title and abstract relevance, (3) supplemented the keyword search through a citation mapping search, (4) screened the results of the citation mapping search for title and abstract relevance, (5) screened the results of both searches for full-text relevance, and (6) collected and synthesized data. 

\subsubsection{Database Search Strategy}
We began with keyword searches in various academic databases. To reach the relevant literature, we considered keywords along two domains: related to body measurement (anthropometry and measurement of `body segment parameters') generally and to motion capture specifically. For both domains, we searched accessible databases containing publications in the biomedical sciences (Scopus, PubMed, Wiley, and Taylor \& Francis). We additionally searched databases focused on the social sciences (JSTOR, Springer) for anthropometry, and databases focused on computer science and electrical engineering (IEEE, ACM) for motion capture. All databases except for PubMed were chosen because they each hosted multiple papers surfaced by a test query (following the query structure in Table \ref{search-strategy}) on Google Scholar. PubMed was then added due to its use in prior motion capture literature reviews  \citep[e.g.,][]{poitras_validity_2019}. Based on the fact that they encompassed the papers surfaced by our test query, we expect that our database selection was robust; however, we additionally included a citation mapping component in order to identify work that may have been excluded. 
\looseness=-1

We restricted our search to results with titles containing words related to error or validity assessment (e.g. “error,” “variability,” “reliability,” etc.) and titles or abstracts referencing the specific measurement domain (e.g. “anthropometry”, “motion capture”, etc.) to ensure that we were only capturing research explicitly attending to error measurement and validation practices within those domains. We conducted our search between March 29th and August 7th, 2023. Databases, search queries, and filters (\S\ref{subsec:filter-search}) are shown in Table \ref{search-strategy}. 

\input{tables/search_strategy}

\subsubsection{Citation Mapping Search Strategy}
To ensure that our keyword search was not overly narrow, we built a citation map based on the keyword search results, which we used to identify additional papers related to error measurement for inclusion. We used ResearchRabbit, \footnote{\url{https://www.researchrabbit.ai/}} a "citation-based literature mapping tool",  to identify similar papers based on overlapping citations. We selected ResearchRabbit over other citation mapping platforms because it (1) allowed for us to build maps starting from one or multiple papers; (2) separated suggested papers into "Earlier," "Similar," and "Later" work, allowing for a clearer understanding of the directionality of citations; and (3) provided full functionality with a free account. However, we note that ResearchRabbit is not without limitations: its documentation is not publicly available, meaning that the specific databases it searches to collect paper metadata and the algorithms it uses to suggest related work are not precisely known. Because our review is focused on the history and foundations of error measurement in motion capture, and because the "Earlier" and "Later" work suggestion algorithms in ResearchRabbit are better-documented \cite{tay_researchrabbit_2021}, we used the “Earlier Work” suggestions to export a set of papers that were cited by our previously surfaced papers. Finally, in cases when work that had been frequently cited by the papers surfaced by our database search was not included in ResearchRabbit (e.g., gray literature like \cite{dempster_space_1955, chandler_investigation_1975}), we manually added it to our set of results.\looseness=-1

\subsubsection{Filtering and Screening Strategy}\label{subsec:filter-search}
After retrieving the raw results of our search query and of our citation mapping search, we restricted our analysis to peer-reviewed articles featuring primary research with full text available online. We further required articles to be written in English because it is the only language in which our entire research team is fluent. Following the application of these filters, one researcher screened the title of each paper for relevance. Relevant titles were those that included keywords related to either measurement of a specific error type or validation of a broader motion capture system. One researcher then screened the abstracts of each remaining paper. Abstracts were considered relevant if they indicated that the paper was specifically about validation or error measurement. Finally, one researcher conducted a full review of the text of each remaining article to determine relevance, again keeping those that were substantively about error measurement for anthropometry or motion capture, or validation of motion capture systems.

\subsubsection{Data Extraction and Analysis}
In our data extraction process, one member of the research team read each paper in full and manually recorded metadata, including whether the paper addressed a specific type of error (and if so, what type) or represented a full-system validation approach, how error or validation was measured, and what data was presented for measurement (if human subjects were used, we recorded their number and demographics). Each of those observations represent details explicitly addressed by the surfaced papers.\looseness=-1

We initially conducted this full review on a sample of fifty papers surfaced through keyword search. Based on this sample, we identified a set of distinct validation and error types within the motion capture literature. We used these error types as keywords to classify the remainder of the papers. If additional common topics appeared as we tagged new papers, we added them to the keyword list as well. For example, we added tags to specify whether another motion capture system was being used as a ``ground truth,'' whether that system was explicitly referred to as a ``gold standard,'' and to indicate when papers proposed systems for use in in-home rehabilitation. We used these keywords to assess trends and standard practices in motion capture measurement and validation. 

While conducting our initial fifty-paper review, we observed that methods for measuring different types of error in motion capture systems were proposed and \textit{stabilized} during distinct periods of time. (We describe these periods, which we call "eras," in detail in \S\ref{sec:findings-errors}.) In other words, \textit{competences} in error measurement were developed shortly after new \textit{materials} for motion capture were introduced, both of which linked up with the \textit{meaning} of motion capture design and innovation. Subsequent work tended to adopt and apply those competences. As a result, although we did not apply date filters in surfacing our original set of papers, we restrict our analysis of each respective error type in \S\ref{sec:findings-errors} to the time period during which its measurement approach was \textit{stabilizing.} 

\subsection{Limitations}
A primary limitation of our method is the risk of bias in both selecting and analyzing the relevant literature. The choice of databases for each error type may have excluded some relevant work hosted on databases other than the ones we searched. We conducted a citation mapping search to mitigate this but we cannot guarantee that it has no impact on our results -- the specific papers suggested by ResearchRabbit also may differ from what a different citation mapping tool may have surfaced, and will still not include research hosted on non-academic sites (e.g. by corporate research \& development teams). The choice to restrict our work to English-language research also excludes valuable global perspectives. Finally, the identities and viewpoints represented in our research team also constitute a source of bias. Although the research team is composed of individuals from a variety of backgrounds, we are all white, cisgender researchers from the United States and Europe. In many ways, our bodies look like the ones that motion capture was designed for. This necessarily hinders our ability to understand the personal impact of motion capture's assumptions about the human body. We also note that none of the authors are primarily motion capture researchers. While this limits the extent to which we are exposed to norms and current practices in the field, it also allows us to examine measurement and validation practice vis-a-vis the history and error typology of motion capture with fresh eyes. Lastly, we want to underscore a conceptual limitation of this paper: whilst assumptions are the graspable thing we are concerned with in this paper, we are not able to cover the siblings of assumptions: politics and ideologies. Due to our focus and scope, the question how these relate to our critical interrogation of assumptions remains unexplored in this paper. This includes, for example, are more focused discussion of how motion capture technology remains an ableist system since it is not only based on a very limited set of data on few homogeneous bodies, but also because it is entangled with ableist ideologies that run across society and technology design at large.

%% file: tables/search_strategy.tex
\begin{table}
\caption[]{Search Strategy. Of note, we included `error' rather than `measurement' in our keyword search because measurement broadly is the domain in which anthropometry and motion capture sit, and the focus on \textit{error measurement} more narrowly allows us to better target work related to implicit assumptions. Similarly, we excluded `variability' and `variation' from our anthropometry-focused keyword search because variation (between individuals or populations) is a standard unit of measurement in anthropometric studies; searches including these keywords returned results that were overly broad. \label{search-strategy}}
\begin{footnotesize}
\centering
\begin{tblr}{
  row{1} = {c},
  cell{2}{4} = {r=2}{},
  hline{1,4} = {-}{0.08em},
  hline{2} = {-}{0.05em},
  hline{3} = {1-3}{0.03em},
}

\textbf{Application Domain} & \textbf{Databases} & \textbf{Search Query} & \textbf{Filters} \\

Anthropometry & 
{JSTOR\\Springer\\Scopus\\PubMed\\Wiley\\Taylor \& Francis} & {\uline{Title} contains:\\ accuracy OR error OR errors OR validity OR \\validation OR validate OR reliability OR \\ consistency OR critical
\newline\\\uline{ Abstract or Title}~contains:\\ anthropometry OR anthropometric OR \\(body AND segment AND parameters)} & 

{Full text accessible online\\Written in English\\Peer-reviewed research\\Full-length paper\\Primary research} \\

Motion Capture & 
{IEEE\\ACM\\Scopus\\PubMed\\Wiley\\Taylor \& Francis} & {\uline{Title} contains:\\ accuracy OR error OR errors OR validity OR \\validation OR validate OR variability OR \\variation OR reliability OR consistency OR critical
\newline\\\uline{ Abstract or Title}~contains:\\ “optical tracking” OR “motion capture” OR \\“motion analysis” OR “motion tracking” OR \\“human motion” OR “3D camera” OR \\stereophotogrammetry} &

\end{tblr}
\end{footnotesize}
\end{table}

%% file: sections/04findings.tex
\input{figures/prisma_results}

\section{Findings: A History of Errors and Assumptions in Motion Capture}
\label{sec:findings-errors}
Following the PRISMA methodology, we analyzed 278 papers related to the measurement in both anthropometry and motion capture technology, as well as validation of motion capture systems (Figure~\ref{prisma-flow}). Our analysis revealed what we categorize as three \textit{eras} in the history of motion capture research: Foundation (1930-1979), Standardization (1980-1999), and Innovation (2000-present). We find that each era centered on a subset of six \emph{error types}: anthropometric errors, errors in body segment parameter estimation, marker placement errors, soft tissue artifacts, instrument errors, and algorithmic errors. Errors make the convergence and stabilization of social practices of measurement and validation legible (Figure \ref{fig:errors-xs}), and are thus an important site for understanding social practices from a historical lens. Although we characterize these eras as discrete historic periods, this does not imply that there are necessarily uniquely distinct, linear eras -- just that these eras highlight key moments during which practices stabilized and different elements were created or became more prevalent. The stabilization and reconfiguration of social practices is messy and contingent; there are moments when practices stabilize, reconfigure, or dissolve as elements shift. These events are always the result of specific histories and contexts.
\looseness=-1

\input{figures/errors-xs}

In each era, technological advances provided new \textit{materials} for measurement and contemporaneous researchers developed new \textit{competences} in making inferences about the human body (recall Figure~\ref{elements_measurement}).
As a result, each era introduces a different set of assumptions underlying the \textit{meanings} of modern motion capture -- and each era measures a different subset of error types corresponding to those assumptions. We propose this historical framework as a tool for the HCI community to understand how motion capture as a social practice has \textit{stabilized} over time \cite{sloane_german_2022}. 
\looseness=-1

\subsection{Foundation Era, (c.\ 1930-1979).}\label{4.2:foundation-era}
The foundations of motion capture technology are based on methods for measuring the human body. Work from the Foundation Era focused extensively on \textit{anthropometry}, i.e., the measurement of bodies (\S\ref{subsubsec:anthropo-errors}). 
A key focus within this was on the \textit{estimation of body segment parameters} (\S\ref{subsubsec:bsp-estimation}). Error measurement during this era was thus correspondingly concerned with these two efforts (Table \ref{tab-errors-foundation}). Already during this era, we begin to see a stabilization around \textit{concurrent validity assessments}: the validation of a new system, tool, or method based on its agreement with a preexisting reference that is considered a `gold standard.' We see a similar stabilization around \textit{reliability assessments}, which quantify the extent to which repeated measurements of the same subject match one another. Both of these remain crucial components of what is understood by `validity' in modern motion capture today. \looseness=-1

\subsubsection{Anthropometric Errors}\label{subsubsec:anthropo-errors}
We observe two main types of anthropometric error of concern in the 9 papers we analyzed from this era: \emph{intra-} and \emph{inter-observer error}. Intra-observer error is defined by the differences in repeated measurements taken by the same observer on the same subject within a short period of time. During this time period, intra-observer error is also referred to as intra-subject error, extrinsic error \cite{davenport_critical_1934}, and reliability \cite{lincoln_reliability_1930, meredith_reliability_1936}. Inter-observer error is defined by the differences in measurements taken by different observers on the same subject.  During the same time period, it was also referred to as precision \cite{marks_reliability_1989}, intrinsic error \cite{davenport_critical_1934}, and objectivity \cite{marshall_objectivity_1937}.

\input{tables/errors_foundation}

Each type of anthropometric error was concerned with repeated measurements of the same subject (i.e., reliability), reflecting attention to the \emph{competences} of anthropometric measurements. In particular, researchers at this time proposed methods for error mitigation (employing multiple measurers \cite{lincoln_reliability_1930} or measuring based on more identifiable bony landmarks \cite{gavan_consistency_1950, martorell_identification_1975}). Some connected \emph{materials} and competences by quantifying inter-observer error from an institutional lens, finding that measurements of the same subjects with the same equipment across two institutions differed systematically \cite{kemper_comparative_1974}. Despite the focus on competences, the historical context indicates that \emph{meanings} were also contingent: noted  eugenicist Charles Davenport, for instance, argued that anthropometry is ``a form of psychometry,'' where these measurements have physiognomic (and essentially racialized) implications \cite{davenport_critical_1934}.

\subsubsection{Errors in Body Segment Parameter Estimation}\label{subsubsec:bsp-estimation}
Motion capture systems use estimated body segment parameters to infer the force required to generate an observed motion -- a crucial component of systems designed, for example, to identify injury risk in athletes and workers. In the Foundation Era, the models through which body segment parameters are estimated---some of which are still in use today---were developed. We analyzed 5 papers from this time. 

\input{figures/body_models}

The earliest influential findings on body segment parameters come from research funded by the US Air Force and therefore was conducted on subjects who were exclusively male, overwhelmingly white, and typically of slender or athletic builds. For instance, Dempster \cite{dempster_space_1955} conducted a study to identify the optimal ergonomic design of a cockpit based on pilots’ range of motion---a classic challenge in interaction design. As part of that work, he acquired, dismembered, and determined the body segment parameters of eight older, white male cadavers. In a similar work, Chandler et al.\ \cite{chandler_investigation_1975} dismembered and measured the BSPs of six male cadavers ``in good overall condition''\footnote{Defined by the authors as without ``congenital anomalies, major surgical alterations, general or localized structural atrophy, excessive wasting, or obesity.''\looseness=-1} in order to better design impact protection systems. Both studies presented simple formulas designed to calculate BSPs based on anthropometric measurements (height; weight; limb length; and, crucially, limb volume) taken of a living body.

The measurements presented by the US Air Force---especially the Dempster study---quickly became accepted as ``ground truth'' for BSPs. This gave rise to the competence of concurrent validity assessment for motion capture: subsequent research used Dempster's data to validate newly proposed BSP formulas. Instead of requiring cumbersome limb volume measurements, these new formulas relied on strong, simplifying assumptions about the human body (specifically, that body segments are rigid and have uniform densities), and then compared their results to Dempster's in order to show that they were valid \cite{hanavan_mathematical_1964, hatze_mathematical_1980}. Meanings are also implicated by this process, which represents an understanding about the kinds of bodies that are and should be measured when estimating human body segment parameters. As such, we can observe a key moment of \emph{stabilization} of the social practices of measurement and validation of BSPs---in this case with respect to concurrent validity and reliance on frozen male cadavers---during this era.\footnote{For a detailed history of body segment parameter estimation practices prior to 1964, see Drillis et al.\ \cite{drillis_body_1964}.} \looseness=-1

\subsection{Standardization Era (c.\ 1980-2000).}\label{subsec:standardization-era} 

The social practices of measurement and validation during the Standardization Era re-stabilized around the development of the first major commercial motion capture systems. These were marker-based motion capture systems, which depend on fixed camera sensors detecting markers attached to specific points on the outside of bodies to infer the associated skeleton and movements (recall \S\ref{subsec:prelim-skeletons}). The key errors attended to during this era were related to the \emph{materials} of these new standardizing technologies, the \emph{meanings} implicit to how they represented and measured bodies, and the new \emph{competences} that they both allowed for and required. Thus we highlight three types of errors: \textit{marker placement errors}, corresponding to competences of the users of these technologies; \textit{soft tissue artifacts}, corresponding to assumptions about the types of bodies motion capture systems are designed for; and \textit{instrument errors}, corresponding to discrepancies associated with the materials themselves (Table \ref{tab-errors-standardization}). In parallel, new measurement and validation strategies were being negotiated as these social practices continued stabilizing, shifting, and re-stabilizing.

\input{tables/errors_standardization}

\subsubsection{Marker Placement Errors}

Making inferences about the movement of a human body with marker-based motion capture systems requires identifying specific locations on a subject's body (with markers), relating them to an anatomical model of a human body, and inferring movement of the anatomical model based on observed movement of the markers. In this context, marker placement error, or anatomical landmark misplacement, refers to the errors from the `misplacement' of markers with respect to certain bone locations; we analyzed 6 papers on this topic from this time. Here, the attention on marker placement errors varies with {palpable} joints (externally identifiable by touching the body, e.g., elbow) or {internal} joints that require inference or imaging to identify (e.g., hip).\footnote{For a detailed history of anatomical landmark misplacement estimation practices prior to 2005, see Della Croce et al.\ \cite{della_croce_human_2005}.}

The anatomical model required to infer movement required assumptions relating the exterior surface of the body to its internal joints. In particular, these assumptions are shaped by meanings around what needs to be measured to define a body -- in other words, anatomical models are adapted to individuals only to the extent to which they can be scaled based on a predefined set of anthropometric measurements \cite{white_three-dimensional_1989}. Whether within or between observers, subjects, or systems, anatomical models were assumed to be stable, with error measurement focused instead on variability due to marker placement, following competences of reliability assessments initially developed for anthropometry (\S\ref{subsubsec:anthropo-errors}) \cite{della_croce_pelvis_1999}. At the same time, a competence emerged of using imaging as a ground truth for marker placements in concurrent validity assessments (following \S\ref{subsubsec:bsp-estimation}) \cite{bell_comparison_1990, neptune_accuracy_1995, leardini_validation_1999}, despite research from as early as 1993 \cite{small_precision_1993} showing that the identification of anatomical landmarks was subjective even in x-rays.\looseness=-1 

\subsubsection{Soft Tissue Artifacts}
In marker-based motion capture systems, markers are placed exterior to the body; in between the markers and the bones of interest are skin, muscle, fat, scar tissue, tendons, viscera, and so on. Markers therefore move independently of bones as subjects’ soft tissues move, introducing both {random} and systematic \textit{soft tissue artifacts} (STA). Systematic errors can emerge from both individual people (e.g., greater error in individuals with more or looser soft tissues due to body type or age) and task type (e.g., greater error in more intense movements).\footnote{For a detailed history of soft tissue artifact estimation practices prior to 2005, see Leardini et al.\ \cite{leardini_human_2005}, or Camomilla et al.\ \cite{camomilla_human_2017} for a more recent editorial.\looseness=-1} The measurement of soft tissue artifacts stabilized in the 1990s; we analyzed 7 papers related to STA measurement published during this time. 
These practices primarily measured STA by comparing skin-mounted markers to bone-mounted (or already implanted \cite{cappozzo_position_1996}) pins \cite{lafortune_three-dimensional_1992,fuller_comparison_1997, holden_surface_1997, reinschmidt_tibiofemoral_1997}, although less invasive imaging methods emerged as well \cite{maslen_radiographic_1994,cappozzo_position_1996}.\looseness=-1

We note several key normative assumptions about the bodies that motion capture is designed for, carried over from the meanings of body measurement highlighted by Dempster's early cadaver studies, that re-emerged during this time. The subjects in foundational STA studies were overwhelmingly male (26 out of 33 subjects for whom sex was specified), able-bodied (all other than the seven subjects recruited due to their bone-mounted surgical pins), and of unremarkable weight (2 out of 37 total subjects were noted to be ``overweight'' according to their BMIs). We again observe how historical assumptions about standard bodies are implicit in these technologies while new elements also take on new logics about bodies (their internal composition and the artifacts of their exteriors, such as loose skin).\looseness=-1

\subsubsection{Instrument Errors}\label{subsubsec:instrument-errors}

The rise of the new materials of marker-based systems additionally led to efforts to measure instrument errors. In this context, \textit{instrument error} refers to inaccuracies in reconstructing the position of marker coordinates as a result of systematic or random errors in the capture of the raw data by the system. Like STA estimation, the competences underlying assessment of instrument error were developed in the 1990s. We identified 7 works measuring instrument error during this time; most emphasize validation methods based on comparisons to rigid objects with known sizes, locations, or movements \cite{haggard_assessing_1990, ehara_comparison_1995, ehara_comparison_1997, klein_accuracy_1995, everaert_measuring_1999, richards_measurement_1999, dabnichki_accuracy_1997}.\footnote{For a detailed review of instrument error measurement practices prior to 2005, see Chiari et al.\ \cite{chiari_human_2005}.} ASTM, a standards organization, named an analogous procedure as the official “Standard Test Method for Evaluating the Performance of Optical Tracking Systems that Measure Six Degrees of Freedom (6DOF) Pose" \cite{e57_committee_test_nodate} in 2016, reflecting a \emph{stabilization} of these practices now codified in technical standards. 
\looseness=-1

\subsection{Innovation Era (c. 2000-present).}\label{subsec:innovation-era}

As sensing technologies became cheaper, faster, and more accessible, and as modern machine learning expanded, we observe another shift in the social practices around measurement and validation in motion capture. Broadened access---through lower cost and less obtrusive systems--diversified the use cases of motion capture systems and the social practices of their design. For instance, consider the shift in materials: alternative sensors like IMUs can collect movement data without expensive camera systems; markerless systems like depth and 2-D video cameras allow unobtrusive surveillance or interaction. At the same time, improved machine learning techniques allow for increased complexity in the underlying body models that motion capture systems rely on to estimate movement and force. Measurement and validation practices in this era focus primarily on measurement of \textit{algorithmic errors} (e.g., of body models), which we discuss below in \S\ref{subsubsec:algo-errors}, and \textit{validation of novel systems} (i.e., asserting that a new technology---which may include new sensors, new processes, and new algorithms---is competent at a historical task), which we discuss in \S\ref{sec:findings-validation}.

\input{tables/errors_innovation}

\subsubsection{Algorithmic Errors}\label{subsubsec:algo-errors}
The development of new materials (sensors, computing capabilities) in motion capture systems corresponds to increased attention on algorithmic errors in body models (competences) used to infer force and motion (Table \ref{tab-errors-innovation}). Specifically, researchers are interested in the deviations between those models and each other, or those models and some external measurement taken as a ground truth. We analyzed 11 papers related to measuring error in body-model related algorithms during the Innovation Era. 

Error in newly proposed body models is often assessed based on their agreement with previous methods (as in the concurrent validity competences used to assess BSP estimation models, discussed in \S\ref{subsubsec:bsp-estimation}) \cite{langley_modified_2021, knodel_electromyography-based_2022, caron-laramee_comparison_2023, moghadam_comparison_2023, myers_validation_2004, bassani_validation_2017}. Concurrent validation may also rely on external sources of data (as opposed to previously existing algorithms using the same data), such as MRIs and EMGs \cite{lai_preliminary_2021, mohtajeb_open_2021}. Building on competences for measuring instrument error (\S\ref{subsubsec:instrument-errors}), some researchers build `dummy' body segments in order to assess the performance of their body models against known quantities of size and movement \cite{sekiguchi_evaluation_2020, myers_validation_2004}. Reliance on benchmark data has also become increasingly common. Researchers increasingly use annotated motion capture datasets like COCO\footnote{\url{https://cocodataset.org/}} or open source motion capture libraries like OpenPose\footnote{\url{https://github.com/CMU-Perceptual-Computing-Lab/openpose}} as ``ground truths'' against which to validate their algorithms \cite{bassani_validation_2017, tsai_enhancing_2022, moghadam_comparison_2023}.\footnote{The design of our systematic review would not recover recent critical scholarship on the politics of training data for computer vision tasks, e.g., \cite{denton2021genealogy,miceli_studying_2022,scheuerman2021datasets}. 
Future attention to motion capture datasets (e.g., in 
\url{http://mocap.cs.cmu.edu/info.php}) in line with existing critiques would, for instance, also reveal a lack of diversity and representations of bodies and their gestures, interactions, and environments.} Others, however, have pointed out that this may negatively impact algorithms' abilities to generalize robustly, as benchmark data may not contain examples of use cases likely to be seen in deployment (a lack of low-light or nighttime images in COCO, for example \cite{chun_nads-net_2019}).\looseness=-1

Another way that has emerged to assess algorithmic error is reformulating the problems that algorithms seek to solve---for example, creating more easily measurable proxies rather than detecting subtle, hard-to-measure body attributes. While this reformulation is standard practice (e.g., \cite{passi_making_2020}), it also reframes the intentions of a system. Fall detection tools, for example, are intended to identify when elderly or gait-impaired individuals exhibit movement patterns that typically precede a fall. However, researchers often train and test models on healthy actors who are asked to walk with predefined abnormal gaits---thereby assuming that these performed gaits are extremely similar to those of individuals with genuine fall risks \cite{chen_computer_2022, wang_approach_2013}. 

Finally, much of the work in the Innovation Era involves leveraging lower cost and less obtrusive sensors to create systems that function well for home use, or are able to work well on populations with atypical body movements (e.g., individuals with Parkinson’s or individuals recovering from surgery or a stroke). Research in this era has primarily focused on the {validation} of these novel systems by comparing their performance to what is that of a `gold standard' marker-based motion capture system, typically in laboratory environments. We explore this practice in detail in \S\ref{sec:findings-validation}.

\subsection{The Evolution of Measurement and Validation Practices in Motion Capture}

\label{subsec:evolution-across-eras}

\input{figures/eras-combined}

Reflecting on the social practices of each era, we can characterize the evolution of the social practices of measurement and validation, as in Figure~\ref{fig:eras-combined}. Considering this evolution, we know that specific elements, i.e., the specific materials, meanings, and competences, shifted across the eras. For instance, calipers (materials) and limb volume calculations (competences) played a significant role in the Foundation Era, whereas benchmark datasets (materials) and calibration (competences) were more typical as elements in the Innovation Era. This evolution can also highlight how these social practices shifted over time. 
\looseness=-1

In the \emph{Foundation Era} (\S\ref{4.2:foundation-era}), motion capture is exclusively evolving around anthropometry, i.e., the measurement of bodies. Validation practices exist, but they are not integral to the practice of motion capture. Error measurement during the Foundation Era is concerned with the estimation of body segment parameters. In the \emph{Standardization Era} (\S\ref{subsec:standardization-era}), practices of measurement and validation being to converge in motion capture, particularly around the meaning of measuring bodies for the design and improvement of motion capture technology. Error measurement during the Standardization Era is concerned with marker placement, soft tissue artifacts, and instrument errors. In the \emph{Innovation Era} (\S\ref{subsec:innovation-era}), practices of measurement and validation are closely converging around the meaning of motion capture generally: both constitute what it means to \emph{do} motion capture. Measurement and validation practices in this era focus primarily on measurement of algorithmic errors and the validation of novel systems.
\looseness=-1

The social practice lens then lets us explicitly compare across eras in Figure \ref{fig:eras-combined}. Here we can visualize this convergence around the meaning of motion capture, as well as how the social practice of validation for motion capture came to be more salient. A more mature practice in the Innovation Era, we next explore the social practices of validation in the Innovation Era in more depth in \S\ref{sec:findings-validation}.
\looseness=-1

\section{Findings: Validation Practices and `Gold Standards' in Motion Capture in the Innovation Era}\label{sec:findings-validation}

\input{tables/validation_methods}
Validation came to play a more central role during the Innovation Era (recall Figure \ref{fig:eras-combined}). Our PRISMA analysis uncovered more variation in the approaches taken to validation in the Innovation Era than in previous eras. Moreover, attention to the \emph{social practice} of validation also helps reveal the origins of this practice via their relationships to elements from earlier eras (Table \ref{tab-validation}). 
\looseness=-1

In the literature from this era, we find a range of approaches. First, we find that \textit{concurrent validity} assessments, in which new technology is compared to a `gold standard,' are considered both necessary and sufficient for validating new motion capture technology, and we attend to how meanings, competences, and practices around `gold standards' come together and shift. \textit{Reliability} assessments, in which new systems are shown to take repeatable measurements, are considered a necessary but not sufficient component of validity. \textit{Construct validity}, which asks whether motion capture systems provide measurements that meaningfully correspond to clinical or other standards, and \textit{robustness}, which measures the ability of a system to maintain performance as its input parameters change, are assessed much less frequently. Crucially, we find that \textit{subgroup validity}, or whether a system performs equally well for individuals with different physical or demographic characteristics, is almost never tested.\looseness=-1

\subsection{Concurrent Validity}
Concurrent validity is by far the most common validation approach our systematic review surfaced, appearing in 97\% (N=228) of the literature. In this approach, a new system or algorithm is compared to a previously validated reference system. Agreement is typically measured through metrics like correlation coefficients or through methods like the Bland-Altman limits of agreement. Although these metrics are simple, they reflect how meanings are operationalized. For example, should a gait parameter be measured as a continuous variable (level of extension of a joint) or a discrete event (peak joint extension)? The choice will impact how accuracy is interpreted and how future practices are shaped.\looseness=-1

We observe stabilizing practices driven primarily by an agreement in materials and meanings: 90\% (N=210) of the time, the reference to which a system is compared is a marker-based motion capture system; 50\% (N=105) of the time such a system is used, it is specifically Vicon-branded.\footnote{https://www.vicon.com/} In 62\% (N=145) of papers, we found that authors explicitly referred to their reference systems as `gold standards.' Once a system has been shown to have even good agreement with an existing `gold standard' through a concurrent validity assessment, it is eligible to be promoted to `gold standard' itself. For example, GAITRite, a walkway outfitted with pressure sensors, was validated against Vicon in 2005 \cite{Webster2005}. Researchers found that GAITRite had a strong, but not perfect, correlation with Vicon in measuring stride time. Twelve years later, a different team used GAITRite to validate the ability of instrumented shoe insoles to measure temporal gait parameters -- and explicitly referred to GAITRite as a ``gold standard" in the process \cite{jagos_mobile_2017}.

A small number of papers we reviewed pointed out that this `gold standard' status carries strong assumptions about what types of error matter. Several research teams argued---correctly---that differences between their proposed systems and the reference systems could have been the result of errors in either system, due to the fact that `gold standard' systems are known to be impacted by marker placement errors and STA \cite{agostini_wearable_2017, potter_evaluation_2022}. The assumptions that remain implicit are also revealing. A common validation approach, for example, involves treating STA in ``gold standard'' systems by treating it as part of the ``ground truth". Rather than attempting to account for STA, researchers fix markers for ``gold standard'' systems directly to the other system's sensors, so that any STA impacts both systems equally \cite{kamstra_quantification_2022}. Agreement between systems---but not the underlying motion of the body---can thus be more easily measured.\looseness=-1

\subsection{Reliability} In 31\% (N=73) of the papers we analyzed, concurrent validity was combined with one or more forms of \textit{reliability} assessment. Most often, this was a \textit{test-retest reliability} assessment, in which researchers confirm that repeated measurements done of the same subject performing the same motion are highly similar to one another (similar to the intra-observer method used to quantify anthropometric errors: \S\ref{subsubsec:anthropo-errors}). 29\% (N=68) of the total papers we examined included such assessments. \textit{Inter-observer reliability}, on the other hand, was undertaken by just 3\% (N=6) of all papers we reviewed.\footnote{We note that we did not consider manual or semi-manual motion analysis systems in our review. These types of systems rely on humans manually annotating landmarks of interest in 2D video data in order to measure kinematic aspects of the human-annotated images. Validation of these systems typically does feature inter-rater reliability metrics (e.g., \cite{herrington_reliability_2017}). However, inter-rater reliability remains an important metric for any system with skin-mounted markers or sensors, even ones that do not rely on human annotation of data. Those systems are still subject to inter-observer variation in marker placements.} For inter-observer reliability, researchers test whether their methods reach similar results when conducted by different measurers (again recalling anthropometric measurement practices). This may entail, for example, calculating the similarity between gait parameters measured by experts in an IMU-based system and clinicians who do not have experience with IMUs \cite{esser_validity_2012}, or examining consistency of measurements taken across different hospitals \cite{cho_evaluation_2018}.\looseness=-1

\subsection{Robustness}
Less common, but still appearing in 9\% (N=21) of all papers, was \textit{robustness}, which is taken to mean the ability of a system to provide consistent measurements as its inputs change; e.g., as cameras move further away from subjects or as time elapses since a system's most recent calibration. To understand whether systems are likely to work across different conditions, researchers examine how well their systems perform with deliberately introduced errors or noise, or with changes in sensor number, orientation, and position \cite{s21196530}. Other studies consider specific challenges likely to be faced in the real world; for example, the robustness of a workplace safety tool to occlusions---something of particular interest for monitoring injury risk in workers who carry large objects \cite{steinebach_accuracy_2020}.\looseness=-1

While these elements of validation are crucial for deploying safe systems, they have not become a standard part of the social practice of validation. A concurrent validity assessment without a robustness component tacitly assumes that a system or technique will continue to be implemented by experts under ideal (laboratory) conditions. However, laboratory setups do not accurately reflect the multitude of environments in which motion capture systems are beginning to be deployed. As these systems are increasingly used in home, clinical, and employment settings, failing to consider robustness may put vulnerable populations at additional risk. \looseness=-1 

\subsection{Construct Validity} Researchers paid mixed attention to \emph{construct validity}, in this context defined as a set of competences to evaluate whether or not a system meaningfully performs its intended use case. We identified construct validity assessments in 15\% (N=34) of the studies we analyzed. This typically involved `{discriminative validity':} is the system able to tell the difference between things that are different \cite{grooten_reliability_2018}? A common version of this discriminative task is to evaluate whether systems can tell the difference between movements conducted by ``healthy" populations and those affected by some disorder that could affect their movements \cite{fanton_validation_2022, jakob_validation_2021, teufl_towards_2019, grooten_reliability_2018, asaeda_validity_2018, aich_validation_2018, summa_validation_2020}. Rather than pure discriminative assessment, some researchers measure whether a system provides clinically relevant information by using it to create scores that can be compared to clinical ratings, like the Foot Posture Index or the Unified Parkinson's Disease Rating Scale \cite{agostinelli_preliminary_2021, tanaka_accuracy_2019, mann_reliability_2014, mcginnis_accuracy_2015, mentiplay_reliability_2013, galna_accuracy_2014}. 

\subsection{User Studies} We identified just seven instances in which researchers conducted \textit{user studies}. Three teams assessed user comfort after multi-hour use of their measurement systems \cite{Buckley2023, auepanwiriyakul_accuracy_2020, OHBERG20211}; one measured response to biofeedback provided by their device as well as comfort \cite{bajpai_novel_2022}. Other approaches included validating that a system to assess athletic performance did not change athletes' behavior \cite{Shepherd2017}, and measuring whether the findings of a system designed to measure muscle stress during lifting tasks aligned with individuals' perceptions of their own levels of effort \cite{hubaut_validation_2022}. The most in-depth user study we analyzed collected feedback from both patients and healthcare providers on the deployment of an interactive remote physical therapy system \cite{Mishra2015}.\looseness=-1 

\subsection{Subgroup Validity} We found a disturbing lack of subgroup validity assessments in motion capture validation -- only 12\% (N=29) of studies considered the relative accuracy of their systems over people of different demographic or physical characteristics. Most of these (N=18) considered the difference in performance of their systems between ``healthy" individuals and those with a specific health status, like Parkinson's disease or having previously had a hip replacement (e.g. \cite{bravi_concurrent_2020, guinet_validity_2021, teufl_towards_2019, jagos_mobile_2017}). Subgroup validation over other kinds of demographic and physical characteristics was much rarer. Four papers considered differential accuracy across participant sex \cite{peebles_validity_2021, hindle_validation_2020, jung_validation_2023, campagnini_estimation_2023}, three across age groups \cite{eltoukhy_validation_2018, auepanwiriyakul_accuracy_2020, Petró_Kiss_2022}, and two across different body types and weights \cite{agostini_wearable_2017, Petró_Kiss_2022}. However, these studies represent the exceptions, and not the rule.\looseness=-1  

Subgroup data was in fact rarely reported at all -- only one of the studies that we identified reported the racial breakdown of their participants (most studies reported height, weight, and sex/gender expressed as a binary variable). That study made a point of deliberately recruiting individuals of varying races and body shapes, although sample sizes were small and they did not explicitly assess performance over those groups \cite{fanton_validation_2022}. In many cases, however, subgroup-specific performance can be hugely important to the potential validity of a tool. A team attempting to develop a device to support low-cost approaches for stroke rehabilitation in Ethiopia, for example, conducted a concurrent validation with a Vicon system in San Francisco. They did not report on the racial breakdown of their subjects (although they did report that 30/31 subjects were right handed), but a host of work on subgroup validity in machine learning (e.g., \cite{buolamwini_gender_2018, wang_towards_2022, banerjee_machine_2021, seyyed-kalantari_chexclusion_2020}) suggests the potential shortcoming of such an approach 
\cite{Hughes2019}.\looseness=-1

The validation of a system on a subject pool that is potentially very different from the population on which it will be deployed has major implications for the system's ability to function in the real world. It also reveals that these practices have and will continue to stabilize around skewed demographics and differential system performance across groups. The few studies that have tested subgroup validity underscore the seriousness of this problem. Durkin and Dowling \cite{durkin_analysis_2003}, for example, examined whether body segment parameters systematically varied across intersectional groups defined by age and gender (and deliberately recruited individuals with diverse bodies within each group). They found significant between-group variation, which remains unaccounted for in most formulas for estimating BSPs \cite{durkin_analysis_2003}. Despite this, the formulas that Dempster developed on eight old, white, male cadavers continue to be treated as a `ground truth' for calculating body segment parameters today, including by systems considered by many to be `gold standards'\footnote{\url{https://docs.vicon.com/display/Nexus215/Plug-in+Gait+output+specification}, archived at \url{https://web.archive.org/web/20230907011629/https://docs.vicon.com/display/Nexus215/Plug-in+Gait+output+specification}}. Researchers have shown that the Dempster-derived formulas do not generalize equally well to individuals who are not older white men \cite{durkin_analysis_2003}, but it is also clear that the formulas do not generalize perfectly to individuals of \emph{any} demographic who are not dead. The cadavers used in these studies had gone through a level of processing (e.g., freezing) that impacted their densities, likely changing their moments of inertia in particular. Standard processes for taking anthropometric measurements of live humans assume that they are standing and breathing. As such, measurements like height or center of gravity might therefore be systematically different between living individuals and cadavers \cite{cardoso_relationship_2016}. 
Recalling how such assumptions are deployed in practice, and that these systems are used in as varied settings as health diagnosis, rehabilitation, and workplace safety surveillance, 
these practices will have real-world impacts, where variation in body types and composition is obscured and and present-day systems optimize for older, white, male (deceased) bodies by design.
\looseness=-1

%% file: figures/prisma_results.tex
\begin{figure*}[t]
\centering
{\includegraphics[width=\textwidth]{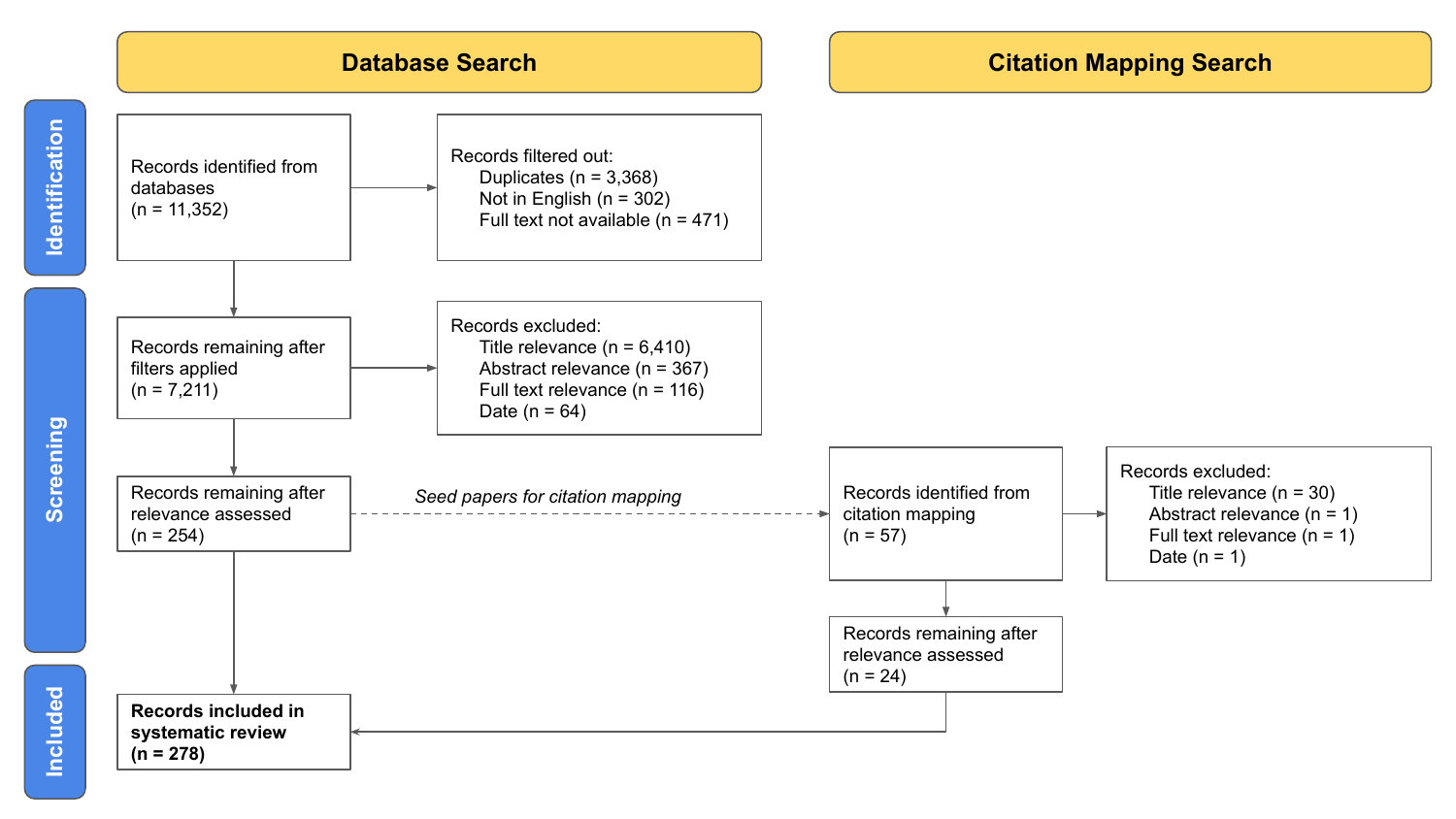}}
\hspace{0.1\textwidth}
\caption{\begin{small} 
PRISMA flow diagram.
\end{small}
\label{prisma-flow}}
\Description{A PRISMA flow diagram showing how many records were identified, screened, and ultimately included in our systematic search. We identified 11,352 papers from databases, filtering out 3,368 as duplicates; 302 non-English papers; and 471 papers without an accessible full text original research work available. This left us with 7,211 papers, after which we applied relevance filters to exclude 6,410 based on title; 367 based on abstract; 116 based on full text; and 64 based on date. This left us with 254 records from our database search. We supplemented that with a citation mapping process, working from our selected papers as seeds. We identified 57 additional records via citation mapping, of which we excluded 30 for title relevance, 1 for abstract relevance, 1 for full text relevance, and 1 for date relevance; leaving us with 24 additional records. The total number of records thus included in our systematic review was 278.}
\end{figure*}

%% file: figures/errors-xs.tex
\begin{figure*}[h]
\centering
{\includegraphics[width=0.9\textwidth]{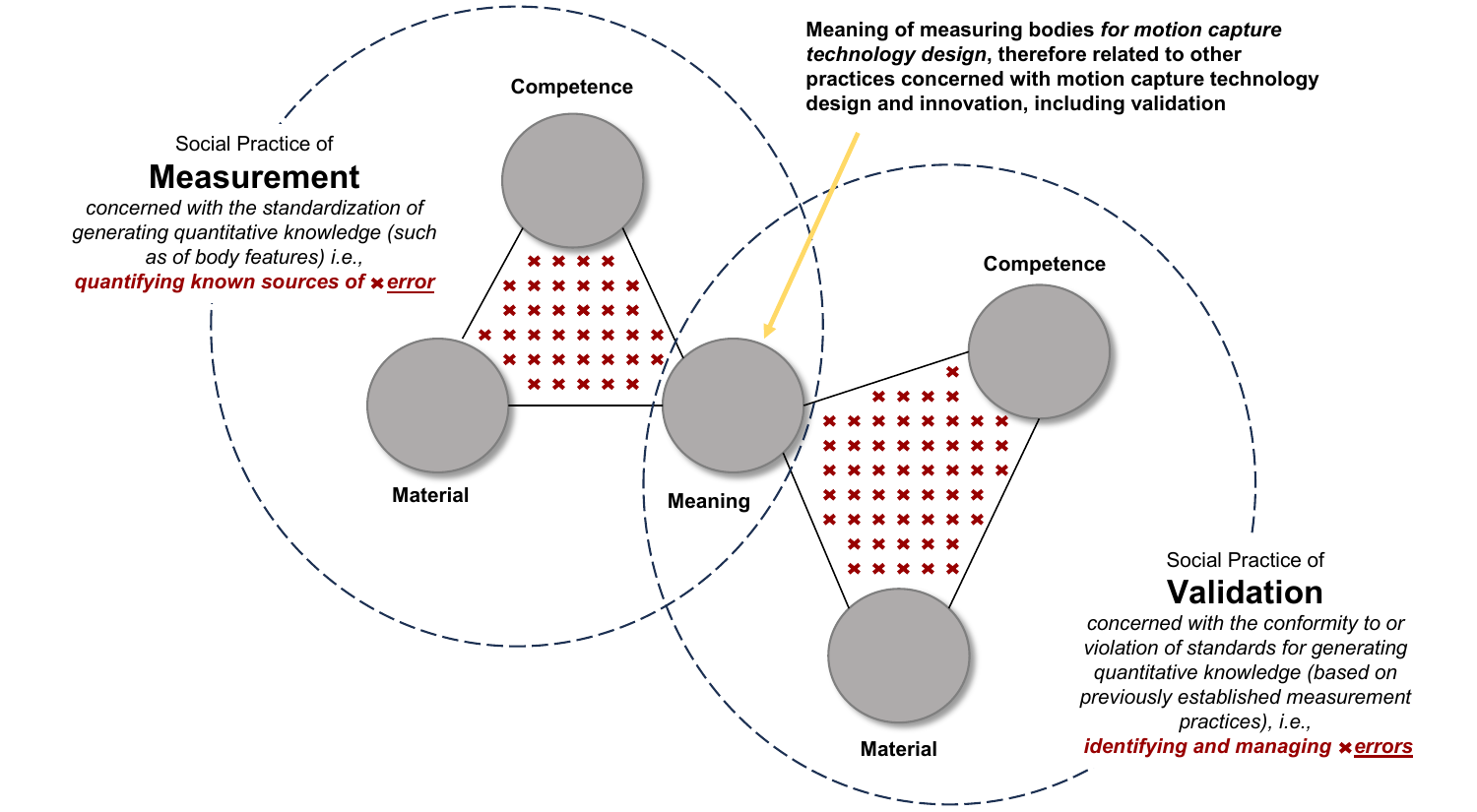}}
\hspace{0.1\textwidth}
\caption{\begin{small} 
The convergence of measurement and validation practices for motion capture.
\looseness=-1
\end{small}
\label{fig:errors-xs}}
\Description{

A diagram that visually represents the convergence of the social practices of measurement and validation within the field of motion capture. On the left, a cluster of three nodes illustrates the 'Social Practice of Measurement', which includes nodes for 'Material', 'Competence', and 'Meaning', all interconnected. The 'Meaning' node extends towards the right, bridging the gap to another cluster of nodes representing the 'Social Practice of Validation', sharing the 'Meaning' node with the measurement practice and connecting to its own 'Material' and 'Competence' nodes. Highlighted within the diagram are numerous red 'X' marks scattered between each of the 'Meaning', 'Competence', 'Material ' node clusters, symbolizing the identification and management of errors within these processes. Accompanying text elaborates on the function of each practice: the left side emphasizes standardization and quantification of known error sources in measurement, while the right side focuses on adherence to or deviation from established standards in validation, particularly concerning error management. The central 'Meaning' node's text underscores its relevance to motion capture technology design and innovation, linking it to broader practices within the field.

}
\end{figure*}

%% file: tables/errors_foundation.tex
\begin{table}
\caption[]{Focus of error measurement in the Foundation Era.\label{tab-errors-foundation}}
\begin{footnotesize}
\centering
\begin{tblr}{
  width = \linewidth,
  colspec = {Q[170]Q[624]Q[146]},
  row{1} = {c},
  hline{1,4} = {-}{0.08em},
  hline{2} = {-}{0.05em},
}
\textbf{Error Type} & \textbf{Description} & \textbf{Papers} \\
Anthropometric Errors & Errors in the measurement (height and weight; length, circumference, mass, and volume of segments) of the human body & \cite{lincoln_reliability_1930, davenport_critical_1934, meredith_reliability_1936, marshall_objectivity_1937, gavan_consistency_1950, munro_analysis_1966, jamison_univariate_1974, kemper_comparative_1974, martorell_identification_1975} \\
Errors in Body Segment Parameter Estimation & Errors in the estimation of the mass, center of gravity, and moment of inertia of individual segments of the human body & \cite{dempster_space_1955, hanavan_mathematical_1964, Clauser1969WeightVA, chandler_investigation_1975, hatze_mathematical_1980}
\end{tblr}
\end{footnotesize}
\end{table}

%% file: figures/body_models.tex
\begin{figure*}
\centering
\subfloat[]{\includegraphics[width=0.3\textwidth]{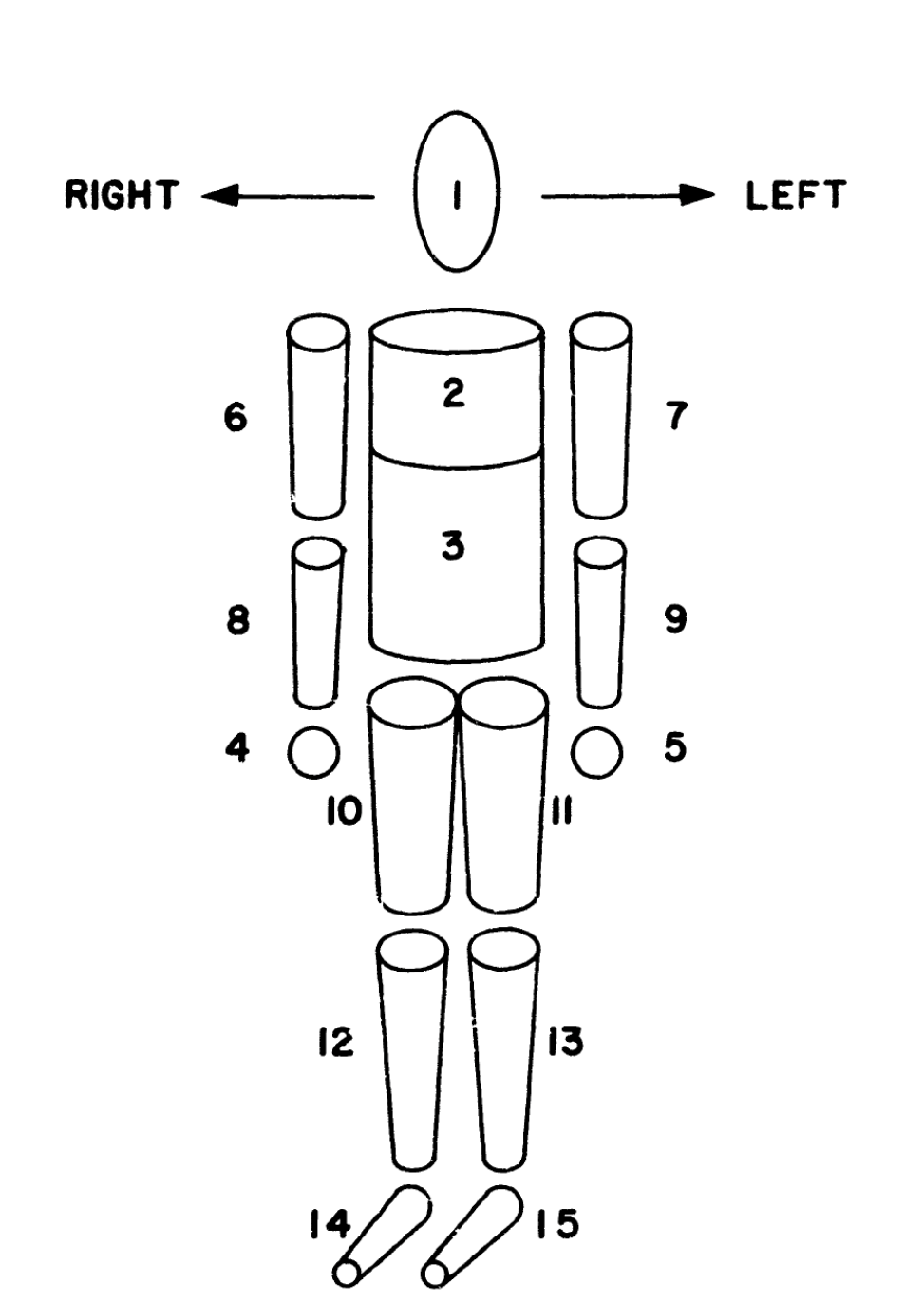}}
\hspace{0.1\textwidth}
\subfloat[]{\includegraphics[width=0.3\textwidth]{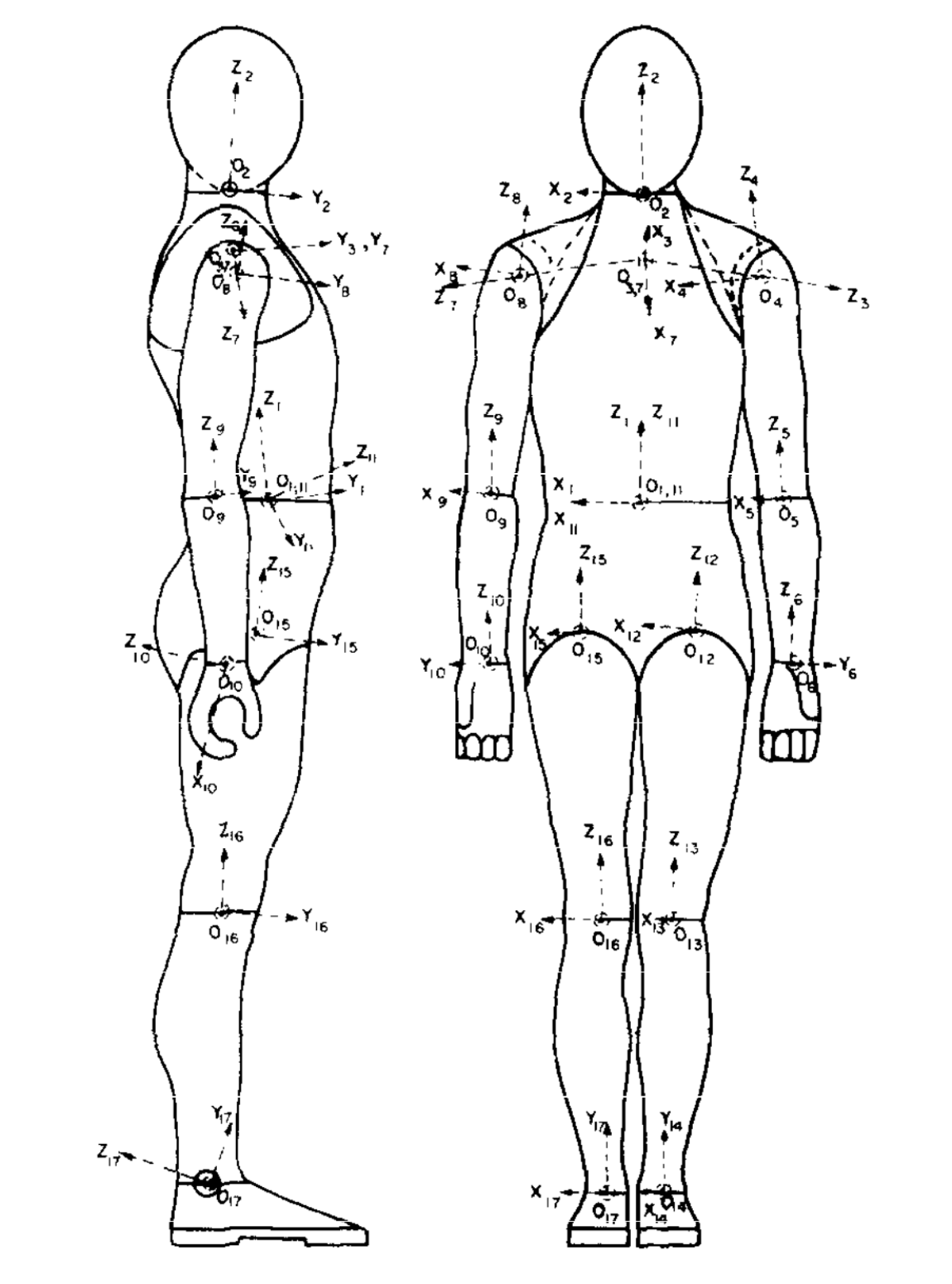}}
\caption{\begin{small} 
Models of the human body used to define and estimate body segment parameters. Although Hanavan's model (a) is more simplistic than Hatze's (b), both rest on crucial assumptions about the rigidity of human body segments that do not reflect reality. Reproduced from Hanavan \cite{hanavan_mathematical_1964} and Hatze \cite{hatze_mathematical_1980}.
\end{small}
\label{body-models}}
\Description{Two models of the human body that have been used to define and estimate body segment parameters. The earlier model on the left is a very simplistic model that divides the body into regular geometric shapes (e.g., cylinder, sphere, cone). The later model on the right is a less simplistic model that looks similar to a mannequin.}
\end{figure*}

%% file: tables/errors_standardization.tex
\begin{table}
\caption[]{Focus of error measurement in the Standardization Era.\label{tab-errors-standardization}}
\begin{footnotesize}
\centering
\begin{tblr}{
  width = \linewidth,
  colspec = {Q[170]Q[580]Q[190]},
  row{1} = {c},
  hline{1,5} = {-}{0.08em},
  hline{2} = {-}{0.05em},
}
\textbf{Error Type} & \textbf{Description} & \textbf{Papers} \\
Marker Placement Errors & Errors in the identification of body landmarks & \cite{white_three-dimensional_1989, bell_comparison_1990, small_precision_1993, neptune_accuracy_1995, della_croce_pelvis_1999, leardini_validation_1999} \\
Soft Tissue Artifacts & Errors in the estimated motion of bones attributable to the fact that the observable surface of the body moves independently of bones & \cite{cappozzo_three-dimensional_1991, lafortune_three-dimensional_1992, maslen_radiographic_1994, cappozzo_position_1996, fuller_comparison_1997, holden_surface_1997, reinschmidt_tibiofemoral_1997} \\
Instrument Errors & Errors in reconstructing 3D coordinates attributable to the capture of raw position data by the system & \cite{haggard_assessing_1990, ehara_comparison_1995, klein_accuracy_1995, dabnichki_accuracy_1997, ehara_comparison_1997, everaert_measuring_1999, richards_measurement_1999}
\end{tblr}
\end{footnotesize}
\end{table}

%% file: tables/errors_innovation.tex
\begin{table}
\caption[]{Focus of error measurement in the Innovation Era.\label{tab-errors-innovation}}
\begin{footnotesize}
\centering
\begin{tblr}{
  width = \linewidth,
  colspec = {Q[140]Q[550]Q[250]},
  row{1} = {c},
  hline{1,3} = {-}{0.08em},
  hline{2} = {-}{0.05em},
}
\textbf{Error Type} & \textbf{Description} & \textbf{Papers} \\
Algorithmic Errors & Errors attributable to the algorithms or models used to analyze captured data & \cite{myers_validation_2004, Wang_novel_2013, bassani_validation_2017, sekiguchi_evaluation_2020, lai_preliminary_2021, mohtajeb_open_2021, langley_modified_2021, tsai_enhancing_2022, knodel_electromyography-based_2022, moghadam_comparison_2023, caron-laramee_comparison_2023} 
\end{tblr}
\end{footnotesize}
\end{table}

%% file: figures/eras-combined.tex
\begin{figure*}[h]
\centering
{\includegraphics[width=0.75\textwidth]{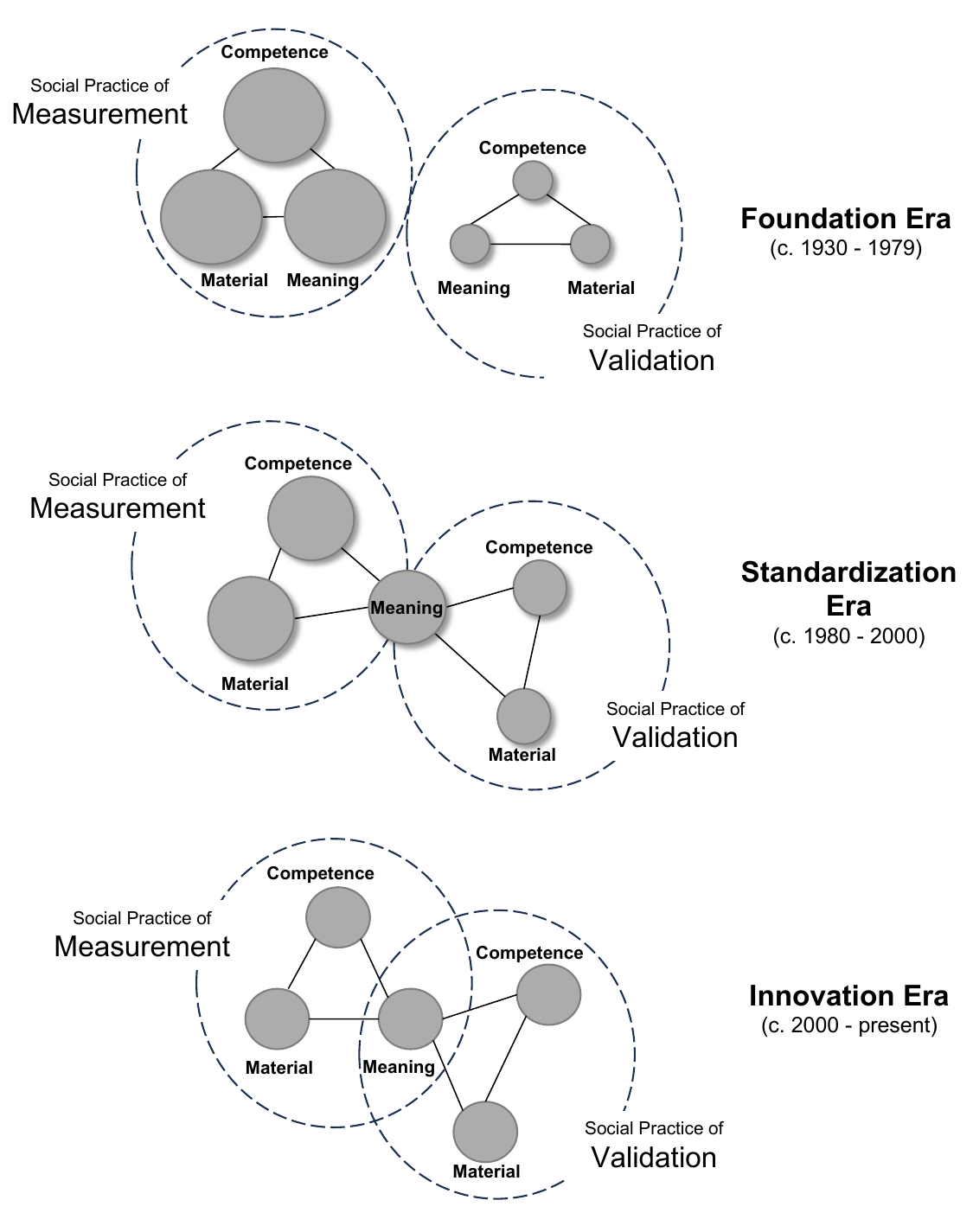}}
\hspace{0.1\textwidth}
\caption{\begin{small} 
Measurement and validation practices converged over the meaning of motion capture of bodies over time. Even though the social practices of measurement and validation were present in each era, both the elements and how they linked up evolved over time. By comparing these diagrams across eras, we highlight how the elements were combined (e.g., converging on meaning) and took on more significant roles (e.g., the elements of validation became more salient over time). \looseness=-1
\end{small}
\label{fig:eras-combined}}
\Description{A series of three sets of diagrams representing the social practice of measurement and validation in different eras. In the 'Foundation Era' at the top, there are two separate structures for 'Social Practice of Measurement' and 'Social Practice of Validation', each with three connected nodes labeled 'Competence', 'Material', and 'Meaning'. The nodes under 'Validation' are smaller than those under 'Measurement'.

In the 'Standardization Era' in the middle, the structures for 'Measurement' and 'Validation' connect at the 'Meaning' node. The 'Competence' and 'Material' nodes in the 'Validation' structure are slightly larger than before but still smaller than those in the 'Measurement' structure.

The 'Innovation Era' diagram at the bottom shows a complex interconnection where the 'Meaning' node links both 'Validation' and 'Measurement' social practices, indicating overlap between the diagrams. Here, the sub-nodes of 'Competence', 'Material', and 'Meaning' are all equal in size. These diagrams visually narrate the historical development and increasing complexity of motion capture practices from 1930 to the present.}
\end{figure*}

%% file: tables/validation_methods.tex
\begin{table}
\caption[]{Motion capture validation practices and their roots.\label{tab-validation}}
\begin{footnotesize}
\centering
\begin{tblr}{
  width = \linewidth,
  colspec = {Q[71]Q[242]Q[627]},
  row{1} = {c},
  hline{1,8} = {-}{0.08em},
  hline{2} = {-}{0.05em},
}
\textbf{Validation Method} & \textbf{Description} & \textbf{What elements of motion capture measurement underlie this validation practice?}\\
\textit{Concurrent Validity} & A new method or system is compared to a previously validated reference~ & Measuring \textbf{errors in body segment parameter estimation }(newly proposed methods were validated through agreement with Dempster) and~\textbf{soft tissue artifacts}~(external sensors were validated by comparison with imaging data or sensors mounted directly on bones)\\
\textit{Reliability} & Repeated measurements of the same subject are compared to one another~ & Measuring \textbf{anthropometric~}and \textbf{marker placement errors}~(intra-observer and inter-observer variation)\\
\textit{Robustness} & System measurements are compared to one another as inputs change~ & Measuring \textbf{instrument errors }(originally to identify ideal conditions for system use) as well as \textbf{soft tissue artifacts }and\textbf{ marker placement errors~}(to identify which marker placements minimize STA)\\
\textit{Construct Validity} & Whether or not a system meaningfully performs its intended use case~ & Measuring \textbf{algorithmic errors}, including assessment of specific \textbf{marker placements }developed for clinical and diagnostic use\\
\textit{User\newline Studies} & Examination of user behaviors and needs & Some roots from original attempts to measure \textbf{body segment parameters }for the purpose of determining space requirements of aircraft operators; also an increasing adoption of HCI methods within motion capture research\\
\textit{Subgroup Validity} & Relative performance of a system over people of different demographic or physical characteristics & An \textbf{interrogation of the assumptions} dating back to the Foundation Era that system performance in general can be well-represented with research on thin, white, male bodies
\end{tblr}
\end{footnotesize}
\end{table}

%% file: sections/06discussion.tex
\section{Discussion}\label{sec:discussion}

The errors we have outlined above correspond to specific, now-hidden assumptions that ground modern instances of motion capture systems and can have major impacts on those who interact with them. Assumptions about the `average' automobile occupant have led to crash tests using dummies approximating a normative male body, which affected automobile safety design in ways that have led to higher injury rates for female occupants \cite{linder_road_2019, perez_invisible_2019}. Similarly, assumptions about the biological basis of race have lead to racial discrimination in technologies used for medical diagnosis \cite{braun_breathing_2014}. Worker safety protocols that make assumptions about certain kinds of bodies and movements create risks of occupational injury for those who do not conform to that model \cite{messing_should_2022}. Each of these cases results---at least in part---from assumptions about subgroup validity. Specifically, when designers assume that a model of an entire population is valid for discrete subpopulations, they place those subpopulations at risk. 

In the case of motion capture, our analysis---conducted through the lens of social practice theory---reveals the dearth of subgroup validity testing within the practice of motion capture. Additionally, it identifies the process through which chains of concurrent validation link even contemporary motion capture systems through the use of `gold standards' to the early estimations of body segment parameters based on only a few white, `athletic,' male cadavers. Our systematic review of how error is treated in the motion capture literature reveals how the elements (materials, competences, and meanings) of measurement and validation practices point to precisely these assumptions and how they are embedded in motion capture systems, in ways that potentially lead to real-world harms.
\looseness=-1

\paragraph{The histories of social practices are crucial to understanding present-day technologies.}

This work offers a {template for the theoretical analysis of social practices}, connecting to existing calls for critical historical perspectives \cite{bodker2023reconnecting}. A social practice theory approach allows us to reveal assumptions---particularly those that assert that a small, privileged group can stand in for the whole---and examine how they have become stabilized and infrastructural as part of technology design. This lens allows us to understand how assumptions achieve significance and permanence in sociotechnical systems. 
\looseness=-1

The context of body measurement, generally, and motion capture, specifically, makes the stakes clear. In the 19th century, Samuel Morton and Louis Agassiz exerted great effort to demonstrate inherent racial differences through measurements of human cranial capacity, skull features, and height \cite{menand_morton_2001, gould_mismeasure_1996}, and Francis Galton attempted to infer character traits like sexuality or criminality from facial features \cite{sekula_body_1986}. Long-since dismissed as pseudo-science, the implications of their work (e.g., racial differences between cranial capacity and intelligence) have been far more durable in public imaginaries of race and bodies -- despite both these measurement practices and their implications also being debunked. This is illustrated by the persistence of arguments about race, cranial capacity, and intelligence \cite{winston_why_2020}, as well as by efforts to revive physiognomy through artificial intelligence \cite{stark_physiognomic_2021}.

Given the dramatically limited efforts to ensure subgroup validity for motion capture technologies identified above, it becomes clear that efforts to apply motion capture to physiognomic goals, whether it be pose estimation or so-called `emotion recognition,' are deeply fraught and unlikely to generalize, even before considering the epistemic bankruptcy of physiognomy. Indeed, a suite of recent research projects dedicated to inferring interior states from motion and pose \cite{rasmussen_using_2023, ahad_towards_2021, lusi_joint_2017, yang_pose-based_2020, kilic_crime_2023} demonstrate both the durability of historical fantasies of physiognomy but also the need to interrogate such projects' underlying assumptions through sociotechnical audits and to address the pernicious \cite{sloane_silicon_2022}, harmful implications of such research \cite{stark_physiognomic_2021}, long before they approach deployment.

\paragraph{Errors point to elements of social practices, assumptions, stakes, and potential interventions}

Attention to the social practices of measurement and validation is of critical importance as motion capture approaches proliferate. However, algorithmic systems that use `gold standard' approaches to validate their outputs propagate the forms of validity held by that gold standard. While the stated motivation of these validation practices are to show that systems work or will be useful, the elements of validation instead stabilize around tasks that are easier to demonstrate. Motion capture is increasingly being developed as marker-free systems that can be deployed at low cost and in public settings. Systems that over- or underestimate the risk of injury in labor-intensive jobs or lead to misdiagnoses of medical conditions, for example, present clear risk of harms; more broadly such designs will likely put already-marginalized populations at greater risk
\cite{bennett2020point,ymous2020just}.\looseness=-1

Identifying populations who are excluded or misrepresented by measurement and validation practices within the context of motion capture can identify contexts where such systems should not yet be deployed without additional testing, or where they should not be deployed at all. This may be particularly relevant for applications where training data is generated by actors portraying medical conditions or physical activities that are not `genuine' instances of these phenomena. While one might assume that actors' portrayals of a condition---for example, one's likelihood of falling (see \S\ref{subsubsec:algo-errors})---are reasonable proxies for `the real thing', these assumptions require interrogation before they are deployed in any use case. A key lens through which to interrogate these assumptions comes from the fields of disability studies and assistive design. The former underscores the significance of lived experience of people with disabilities and their intersectional identities in technology design \cite{schalk_black_2022, williamson_accessible_2019, hamraie_building_2017}, pointing out that people with disabilities are often already technology designers \cite{hamraie_crip_2019}. The latter relies on user-centered design principles that emphasize the importance of user involvement in the design process -- in other words, making clear that ``designers cannot play the role of disabled users, even if they try to recreate their conditions of life'' \cite{thomann_designing_2016}.\looseness=-1

\paragraph{Social practices yield key insights into the design of systems and the study of interactions}
It is well established in HCI research that technologies and systems change human behavior and vice versa \cite{klemmerHowBodiesMatter2006,coeckelbergh2019moved}. As we interact with our social world through technologies, we continually adapt and comport ourselves differently with that knowledge in mind -- and this affects how we, quite literally, move through the world. For example, a person who switches from an analog to a smartwatch is likely to adapt their `watch-checking motion' in order to activate the smartwatch screen. Social practice theory emphasizes exploring how that change in human-machine interaction occurs and becomes a stable phenomenon that we can observe in the world. The premise is that interactions and behaviors change first the elements of individual practices, and then social practices at large. Rather than exploring just (technology design) intentions on the one hand and (technology design) impacts on the other, the study of the dynamics of technology design and interactions moves decidedly into focus, and with it, a view into the assumptions that underpin them. As our analysis of the assumptions underpinning measurement and validation practice in motion capture technology has shown, assumptions can become solidified as they are baked into the material nature of a technology practice. This conceptual approach is particularly useful for understanding how `bias' enters hardware-driven systems---such as sensor-centric technologies like motion capture, volumetric technology, or other camera-based systems---and becomes an infrastructural condition.

\paragraph{Accountable technologies require understanding social practices}

In interrogating social practices of measurement and validation, the assumptions about what constitutes measurement, error, and validity can contribute to socio-technical audits of technology to ensure it has adequate epistemological grounding \cite{sloane_silicon_2022, rhea_external_2022} for its context of use.

To this end, understanding the social practices of measurement and validation for motion capture are necessary precursors for developing a sociotechnical audit of motion capture systems \cite{sloane_silicon_2022}. Understanding assumptions made about bodies by motion capture systems (the measurements needed to define them, how their skeletons and soft tissues move, how they vary systematically across a population) can aid the design of such audits, directing attention toward whether the use of golden standards in validating a system was justified or appropriate and whether motion capture systems perform as expected for subjects of varying body types, physical abilities, demographics, and so on (i.e., whether the system displays subgroup validity). Fleshing out assumptions is the basis for more holistic AI audit approaches that push beyond questions of disparate impact \cite{sloane_algorithmic_2021} and that center on the question of a system's stability and its status of a valid measurement instrument \cite{sloane_silicon_2022, rhea_external_2022, sloane_make_2022, ma_you_2023}. In other words, focusing on assumptions allows for AI audits that are not purely technical in nature, but that examine whether an AI  functions in alignment with its intended purpose while simultaneously inspecting the validity of the principles underpinning its design \cite{sloane_silicon_2022,jacobs_measurement_2021}.

Many data-driven technologies other than motion capture incorporate their own set of social practices for measurement and validation. The work presented here provides a model through which these social practices can be interrogated, historically and ethnographically. Researchers and technologists who wish to use the social practice theory lens for the \textit{design} or \textit{assessment} of technologies can start off with the following questions: What are the key social practices comprising the sociotechnical system at play (for example, algorithmic risk assessments or facial recognition technology)? What are the materials, competences, and meanings of these practices? As these elements link up and stabilize the practice, what assumptions are simultaneously stabilized? Where do these assumptions come from and they potentially harmful? The visuals provided in this paper may be of additional help for this exercise. Peeling back the layers of social practices in technology in this way cannot only help with executing more holistic AI audits, it can also, for example, help to systematically assess AI ethics interventions \cite{sloane_german_2022}  or find effective ways of complying with AI regulation \cite{sloane2022introducing}. 
\looseness=-1

\section{Conclusion}\label{sec:conclusion}

We offer a template for the theoretical analysis of social practice in the design and use of technologies. By focusing on the social practices of measurement and validation for motion capture, we connect with broader histories of social and technical representations of bodies while shedding light on the tenuous foundations of modern systems. We uncover key stabilizing moments of measurement and validation across three historical eras and propose a typology of errors, revealing shifting attention and inculcation of past elements through validation practices that entrench, rather than challenge, them. This type of analysis can reveal how assumptions built into technologies shift, stabilize, and become infrastructural in the design and use of technologies. 
\looseness=-1